\documentclass[12pt]{article}

\def\beq{\begin{equation}}
\def\endeq{\end{equation}}
\def\beqn{\begin{eqnarray}}
\def\endeqn{\end{eqnarray}}

\newcommand{\cJ}{{\cal J}}
\newcommand{\cK}{{\cal K}}
\newcommand{\cL}{{\cal L}}

\newcommand{\cS}{{\cal S}}

\input{tcilatex}

\begin{document}

\title{Measurements continuous in time and a posteriori states in quantum
mechanics}
\author{A Barchielli \\
Dipartimento di Fisica dell'Universit\`{a} di Milano,\\
and Istituto Nazionale di Fisica Nucleare,\\
Sezione di Milano, Utaly \and V P Belavkin \\
MIEM, Moscow 109028, USSR}
\date{30 May 1990\\
Published in: J. Phys. A: Math. Gen. \textbf{24} (1991) 1495--1514 }
\maketitle

\begin{abstract}
Measurements continuous in time were consistently introduced in quantum
mechanics and applications worked out, mainly in quantum optics. In this
context a quantum filtering theory has been developed giving the reduced
state after the measurement when a certain trajectory of the measured
observables is registered (the a posteriori states). In this paper a new
derivation of filtering equations is presented, in the cases of counting
processes and of measurement processes of diffusive type. It is also shown
that the equation for the a posteriori dynamics in the diffusive case can be
obtained, by a suitable limit, from that one in the counting case. Moreover,
the paper is intended to clarify the meaning of the various concepts
involved and to discuss the connections among them. As an illustration of
the theory, simple models are worked out.
\end{abstract}


\nocite{bib:bhaml2,bib:bmes2,bib:bmes3,bib:bmes4,bib:bmes5,bib:bmes6} \nocite%
{bib:bmes7,bib:bmes8,bib:bmes9,bib:bhaml3,bib:bmes11} \nocite%
{bib:bmes12,bib:bel1,bib:bmes14,bib:bmes15,bib:bmes16,bib:bmes17} \nocite%
{bib:b5,bib:bmes19,bib:XP26,bib:ref2,bib:XP31,bib:XP38} \nocite%
{bib:ref11,bib:ref13,bib:ref14,bib:ref17,bib:ref19,bib:bel4} \nocite%
{bib:bmes30,bib:bmes31,bib:bmes32,bib:bmes33,bib:bmes34} \nocite%
{bib:bmes35,bib:bmes36,bib:bmes37,bib:ref26,bib:ref21} \nocite%
{bib:bmes40,bib:bhaml17,bib:bmes42,bib:bmes43,bib:ref25} \nocite%
{bib:bel8,bib:bmes46,bib:bhaml18,bib:bmes48,bib:bmes49,bib:bel10} \nocite%
{bib:bmes51,bib:bmes52,bib:ref18}



\section*{Introduction}

\typeout{Introduction} \label{sec:bmes0} \setcounter{equation}{0}

Usually in quantum mechanics only instantaneous measurements are considered,
but by using the notion of \emph{instrument\/} \cite{bib:bhaml2}--\cite%
{bib:bmes3} also measurements continuous in time were consistently
introduced \cite{bib:bmes2,bib:bmes4}--\cite{bib:bmes15} and applications
worked out \cite{bib:bmes2,bib:bmes7,bib:bmes9,bib:bmes12,bib:bmes16}--\cite%
{bib:bmes19}; see also \cite{bib:XP26,bib:XP38}.

Now a natural question is: if during a continuous measurement a certain
trajectory of the measured observable is registered, what is the state of
the system soon after, conditioned upon this information (the ``a
posteriori" state)? By using ideas from the classical filtering theory for
stochastic processes and the formulation of continuous measurements in terms
of quantum stochastic differential equations \cite%
{bib:bel1,bib:bmes14,bib:bmes17,bib:XP31,bib:ref11}, a stochastic equation
for the a posteriori states has been obtained \cite{bib:ref13}--\cite%
{bib:ref19}. The main purpose of this paper is indeed that of clarifying the
meaning of that equation by presenting a natural derivation of it in terms
of instruments, independently from any notion related to quantum stochastic
calculus, and by discussing some models. Moreover, we shall discuss the
connections among various things appeared in the literature about what can
be called a quantum version of the theory of stochastic processes
(continuous measurements) and filtering theory (a posteriori states).

Let us start by recalling the important notions of instrument and of a
posteriori states. The notion of instrument has been introduced in the
operational approach to quantum mechanics \cite{bib:bhaml2}. Let a quantum
system be described in a separable Hilbert space ${\mathcal{H}}$ and denote
by ${\mathcal{B}}({\mathcal{H}})$ and ${\mathcal{T}}({\mathcal{H}})$ the
Banach spaces of the bounded operators on ${\mathcal{H}}$ and the
trace--class operators, respectively. Let $(\Omega ,\Sigma )$ be a
measurable space ($\Omega $ a set and $\Sigma $ a $\sigma $-algebra of
subsets of $\Omega $). An \emph{instrument} \cite{bib:bhaml2}--\cite%
{bib:bmes3} $\mathcal{I}$ is a map from $\Sigma $ into the space of the 
\emph{linear} bounded operators on ${\mathcal{T}}({\mathcal{H}})$ such that
(i) $\mathcal{I}(B)$ is completely positive \cite{bib:bel4} for any $B\in
\Sigma $, (ii) $\sum_{j}\mathcal{I}(B_{j})\varrho =\mathcal{I}\big(%
\bigcup_{j}B_{j}\big)\varrho \ $ for any sequence of pairwise disjoint
elements of $\Sigma $ and any $\varrho $ in ${\mathcal{T}}({\mathcal{H}})$
(convergence in trace norm), (iii) $\mathrm{Tr}\{\mathcal{I}(\Omega )\varrho
\}=\mathrm{Tr}\{\varrho \}$,\quad $\forall \varrho \in \mathcal{T}(\mathcal{H%
})$.

The instrument $\mathcal{I}$ is an operator--valued measure: (i) is the
positivity condition, (ii) is $\sigma $-additivity, (iii) is normalization.
The instruments represent measurement procedures and their interpretation is
as follows. $\Omega $ is the set of all possible outcomes of the measurement 
$\big((\Omega ,\Sigma )$ is called the value space$\big)$ and the
probability of obtaining the result $\omega \in B$ $(B\in \Sigma )$, when
before the measurement the system is in a state $\varrho $ $\big(\varrho \in 
{\mathcal{T}}({\mathcal{H}})$, $\varrho \geq 0$, $\mathrm{Tr}\{\varrho \}=1%
\big)$, is given by $P(B|\varrho ):=\mathrm{Tr}\{\mathcal{I}(B)\varrho \}$.
Moreover, let us consider a sequence of measurements represented by the
instruments $\mathcal{I}_{1}$, $\mathcal{I}_{2}$, ..., $\mathcal{I}_{n}$ and
performed in the natural order ($\mathcal{I}_{2}$ after $\mathcal{I}_{1}$
and so on). We assume any time specification to be included in the
definition of the instruments (Heisenberg picture). Then, the joint
probability of the sequence of results $\omega _{1}\in B_{1}$, $\omega
_{2}\in B_{2}$, ..., $\omega _{n}\in B_{n}$, when the \emph{premeasurement
state} is $\varrho $, is given by 
\begin{equation}
P(B_{1},B_{2},\ldots ,B_{n}|\varrho )=\mathrm{Tr}\left\{ \mathcal{I}%
_{n}(B_{n})\circ \mathcal{I}_{n-1}(B_{n-1})\circ \cdots \circ \mathcal{I}%
_{1}(B_{1})\varrho \right\} \,.  \label{eq:1.0a}
\end{equation}%
If we consider the conditional probability of the results $\omega _{2}\in
B_{2}$, ..., $\omega _{n}\in B_{n}$ given the first result $\omega _{1}\in
B_{1}$, we can write 
\begin{eqnarray}
P(B_{2},\ldots ,B_{n}|B_{1};\varrho ) &\equiv &{\frac{P(B_{1},B_{2},\ldots
,B_{n}|\varrho )}{P(B_{1}|\varrho )}}=  \nonumber \\
=P\big(B_{2},\ldots ,B_{n}|\varrho (B_{1})\big) &\equiv &\mathrm{Tr}\left\{ 
\mathcal{I}_{n}(B_{n})\circ \cdots \circ \mathcal{I}_{2}(B_{2})\varrho
(B_{1})\right\} \,,  \label{eq:1.0b}
\end{eqnarray}%
where we have introduced the statistical operator $\varrho (B_{1})$
representing the state after the first measurement, conditioned upon the
result $\omega _{1}\in B_{1}$. For a generic instrument $\mathcal{I}$ and
set $B$, the conditioned state $\varrho (B)$ is defined by 
\begin{equation}
\varrho (B)={\frac{\mathcal{I}(B)\varrho }{\mathrm{Tr}\{\mathcal{I}%
(B)\varrho \}}}\equiv {\frac{\mathcal{I}(B)\varrho }{P(B|\varrho )}}\,.
\label{eq:1.0c}
\end{equation}%
Let us note that joint probabilities (\ref{eq:1.0a}) preserve mixtures, by
the linearity of the instruments: for $\varrho $ and $\sigma $ statistical
operators and $0\leq \lambda \leq 1$, we have 
\begin{equation}
\lambda P(B_{1},\ldots ,B_{n}|\varrho )+(1-\lambda )P(B_{1},\ldots
,B_{n}|\sigma )=P\big(B_{1},\ldots ,B_{n}\big|\lambda \varrho +(1-\lambda
)\sigma \big)\,.  \label{eq:1.0d}
\end{equation}%
However, this property is not shared by conditional probabilities (\ref%
{eq:1.0b}), by the definition itself of conditioning, and, therefore, the
expression (\ref{eq:1.0c}) for the conditioned state is not linear in the
premeasurement state $\varrho $, unless $B=\Omega $.

Formula (\ref{eq:1.0c}) can be interpreted by saying that we perform some
measurement on a statistical ensemble of systems and select those systems
for which the result $\omega\in B$ has been found. Then, (\ref{eq:1.0c}) is
the state after the measurement of the systems selected in this way and
depends not only on the result $\omega\in B$, but also on the perturbations
due to the concrete measuring procedure and to the dynamics. If we perform
the measurement, but no selection, we obtain by (iii) $\varrho(\Omega) = 
\mathcal{I}(\Omega)\varrho$. By the definition of instrument, this quantity
is linear in $\varrho$ and it is a statistical operator if $\varrho$ is a
state. We can call $\varrho(\Omega)$ the \emph{a priori state}: if we know
the premeasurement state $\varrho$ and the measurement $\mathcal{I}$, $%
\varrho(\Omega)$ is the state we can ``a priori" attribute to our systems,
before knowing the result of the measurement.

Let us consider now the case of the most fine selection when in (\ref%
{eq:1.0c}) the set $B$ shrinks to an \textquotedblleft infinitesimally
small" set $\mathrm{d}\omega $ around the value $\omega $. According to the
discussion above, the quantity 
\begin{equation}
\varrho (\omega )={\frac{\mathcal{I}(d\omega )\varrho }{\mathrm{Tr}\{%
\mathcal{I}(d\omega )\varrho \}}}  \label{eq:1.1}
\end{equation}%
represents the state conditioned upon the result $\omega \in \mathrm{d}%
\omega $. The quantity $\varrho (\omega )$ is the state one can attribute to
those systems for which the result $\omega $ has actually been found in the
measurement and for this reason we call it a posteriori state \cite{bib:XP38}%
.

More precisely, a family of statistical operators $\{\varrho (\omega )$, $%
\omega \in \Omega \}$ is said to be a family of \emph{a posteriori states} 
\cite{bib:bmes30}, for an initial state $\varrho $ and an instrument $%
\mathcal{I}$ with value space $(\Omega ,\Sigma )$, if (a) the function $%
\omega \rightarrow \varrho (\omega )$ is strongly measurable with respect to
the probability measure 
\begin{equation}
\mu _{\varrho }(B):=\mathrm{Tr}\{\mathcal{I}(B)\varrho \}\equiv P(B|\varrho )
\label{eq:1.3}
\end{equation}%
for the observable associated with the instrument $\mathcal{I}$ and (b) $%
\forall Y\in {\mathcal{B}}({\mathcal{H}})$, $\forall B\in \Sigma $, 
\begin{equation}
\int_{B}\mathrm{Tr}\{Y\varrho (\omega )\}\mu _{\varrho }(\mathrm{d}\omega )=%
\mathrm{Tr}\{Y\,\mathcal{I}(B)\varrho \}\,.  \label{eq:1.2}
\end{equation}%
Let us note that by definition the link between a priori and a posteriori
states is given by 
\begin{equation}
\varrho (\Omega )\equiv \mathcal{I}(\Omega )\varrho =\int_{\Omega }\varrho
(\omega )\,\mu _{\varrho }(\mathrm{d}\omega )\,.  \label{eq:1.4}
\end{equation}

Let us stress that (\ref{eq:1.2}) defines the a posteriori states once the
instrument $\mathcal{I}$ and the \emph{premeasurement state} $\varrho$ are
given. On the contrary, if $\varrho(\omega)$ and $\mu_{\varrho}$ are given
for any $\varrho$, (\ref{eq:1.2}) allows to reconstruct the instrument $%
\mathcal{I}$. We shall make use of this in the following sections.

Finally, let us note that there is no reason for $\varrho(\omega)$ to be a
pure state if $\varrho$ is pure: it depends on the concrete measuring
procedure. Roughly speaking $\varrho(\omega)$ is pure if one has some
property of minimal disturbance, some ability of the measurement to give a
maximum of information, ...; we shall see various examples ($\varrho(\omega)$
pure and not pure) in the case of continuous measurements.

\section{Counting processes}

\typeout{Counting processes} \label{sec:bmes1} \setcounter{equation}{0}

The first class of continuous measurements which has been introduced in
quantum mechanics is that of counting processes \cite{bib:bmes2,bib:bmes4}--%
\cite{bib:bmes9,bib:bel1,bib:bmes17,bib:b5,bib:ref19,bib:bmes31}-- \cite%
{bib:bmes34}. One or more counters act continuously on the system and
register the times of arrival of photons or other kinds of particles.

Let us consider the case of $d$ counters. They differ by their localization
and/or by the type of particles to which they are sensible and/or by their
operating way... We can describe this counting process by giving the so
called \emph{exclusive probability densities\/} (EPDs) \cite%
{bib:bmes7,bib:bmes9}. The quantity $P_{t_{0}}^{t}(0|\varrho)$ is the
probability of having no count in the time interval $(t_{0},t]$, when the
system is prepared in the state $\varrho$ at time $t_{0}$. The quantity $%
p_{t_{0}}^{t}(j_{1},t_{1};j_{2},t_{2};\ldots;j_{m},t_{m}|\varrho)$, $j_k =
1, \ldots, d$, $t_0 < t_1 < t_2 < \cdots < t_m \leq t,$ is the multi--time
probability density of having a count of type $j_1$ at time $t_1$, a count
of type $j_2$ at time $t_2$, $\ldots$, and no other count in the rest of the
interval $(t_0,t]$. Davies \cite{bib:bmes4} (see also \cite{bib:bmes5}--\cite%
{bib:bmes9}) has shown that these EPDs can be consistently described in
quantum mechanics in the following way.

Let $\mathcal{L}_{0}(t)$ be a Liouvillian $\big($the generator of a
completely positive dynamics \cite{bib:bel4} on ${\mathcal{T}}(\mathcal{H})%
\big)$ and $\mathcal{J}_{j}(t)$, $j=1,\ldots ,d$, be completely positive
maps on ${\mathcal{T}}({\mathcal{H}})$. Let us introduce the positive
operators $R_{j}(t)$ on ${\mathcal{H}}$ by 
\begin{equation}
R_{j}(t)\!:\,=\mathcal{J}_{j}(t)^{\prime }\mathbf{1}\,.  \label{eq:2.1}
\end{equation}%
For any operation ${\mathcal{A}}$ on ${\mathcal{T}}({\mathcal{H}})$ its
adjoint ${\mathcal{A}}^{\prime }$ on ${\mathcal{B}}({\mathcal{H}})$ is
defined by 
\begin{equation}
\mathrm{Tr}\{X\,{\mathcal{A}}\varrho \}=\mathrm{Tr}\{\varrho \,{\mathcal{A}}%
^{\prime }X\}\,,\qquad \forall \varrho \in {\mathcal{T}}({\mathcal{H}}%
)\,,\quad \forall X\in {\mathcal{B}}({\mathcal{H}})\,.  \label{eq:2.2}
\end{equation}

Finally, let $\mathcal{S}(t,t_{0})$, $t\geq t_{0}$, be the family of
completely positive maps on ${\mathcal{T}}({\mathcal{H}})$ defined by the
equations 
\begin{equation}
{\frac{\partial }{\partial t}}\mathcal{S}(t,t_{0})=\mathcal{A}(t)\mathcal{S}%
(t,t_{0})\,,\qquad \mathcal{S}(t_{0},t_{0})=\mathrm{Id}\,,  \label{eq:2.3}
\end{equation}%
\begin{equation}
\mathcal{A}(t)\varrho =\mathcal{L}_{0}(t)\varrho -{\frac{1}{2}}%
\sum_{j=1}^{d}\{R_{j}(t),\varrho \}\,.  \label{eq:2.4}
\end{equation}%
Here $\{a,b\}=ab+ba$ and $\mathrm{Id}$ is the identity map on $\mathcal{T}(%
\mathcal{H})$. Then, the quantities 
\begin{equation}
P_{t_{0}}^{t}(0|\varrho )=\mathrm{Tr}\{\mathcal{S}(t,t_{0})\varrho \}\,,
\label{eq:2.5}
\end{equation}%
\begin{eqnarray}
\lefteqn{p_{t_{0}}^{t}(j_{1},t_{1};j_{2},t_{2};\ldots ;j_{m},t_{m}|\varrho )=%
\mathrm{Tr}\Bigl\{\cS(t,t_{m})\cJ_{j_{m}}(t_{m})}  \nonumber \\
&&\times \mathcal{S}(t_{m},t_{m-1})\mathcal{J}_{j_{m-1}}(t_{m-1})\cdots 
\mathcal{S}(t_{2},t_{1})\mathcal{J}_{j_{1}}(t_{1})\mathcal{S}%
(t_{1},t_{0})\varrho \Bigr\}  \label{eq:2.6}
\end{eqnarray}%
(where $t_{0}<t_{1}<t_{2}<\cdots <t_{m}\leq t$, $j_{k}=1,\ldots ,d$) are a
consistent family of EPDs.

The whole statistics of the counts can be reconstructed from the EPDs. For
instance, the probability of $m$ counts of type $j$ in the time interval $%
(t_0,t_1]$, $n$ counts of type $i$ in the interval $(t_1,t_2]$ is given by 
\begin{eqnarray}
\lefteqn{P\big(m,j,(t_0,t_1];n,i,(t_1,t_2]|\varrho\big) = }  \nonumber \\
& &=\int_{t_1}^{t_2}\mathrm{d} r_n \int_{t_1}^{r_n}\mathrm{d} r_{n-1} \cdots
\int_{t_1}^{r_2}\mathrm{d} r_1 \int_{t_0}^{t_1}\d s_m \int_{t_0}^{s_m}%
\mathrm{d} s_{m-1} \cdots  \nonumber \\
& &\cdots \int_{t_0}^{s_2}\mathrm{d} s_1 \ p_{t_0}^{t_2}\big(%
j,s_1;j,s_2;\ldots;j,s_m;i,r_1;i,r_2;\ldots;i,r_n| \varrho\big)\, .
\label{eq:2.7}
\end{eqnarray}
In a similar way all more complicated joint probabilities can be constructed.

One of the most significant problems treated by this theory is that of the
electron shelving effect or quantum jumps. An atom with a peculiar level
configuration and suitably stimulated by laser light emits a pulsed
fluorescence light with random bright and dark periods. It is possible to
use $\mathcal{L}_0(t)$ for describing the free atom and the driving term due
to the laser and to use the operators $\mathcal{J}_j(t)$ for describing the
emission process. Then, the full statistics of the fluorescence light can be
computed and, in particular, the mean duration of the dark periods \cite%
{bib:bmes17}--\cite{bib:bmes19}. Other applications to quantum optics of the
counting theory described here are given in \cite{bib:bmes35}--\cite%
{bib:bmes37}.

Let us now consider the problem of the a posteriori states. Our counting
process can be considered as a stochastic process whose associated
probability measure $\big($uniquely determined by (\ref{eq:2.5}) and (\ref%
{eq:2.6})$\big)$ is concentrated on step functions. Let us consider $t_{0}=0$
as initial time. A typical trajectory $\omega _{t}$ up to time $t$ is
specified by giving the sequence $(j_{1},t_{1};j_{2},t_{2};\ldots
;j_{n},t_{n})$ of types of counts and instants of counts $t_{1}<t_{2}<\cdots
<t_{n}$ up to time $t$. Let $\omega _{t}$ be the trajectory we have
registered up to time $t$. Then, the conditional probability $P\big(%
0,(t,t\!+\!\overline{t}]|\omega _{t};\varrho \big)$ of no count in the
interval $(t,t\!+\!\overline{t}]$, given the state $\varrho $ at time zero
and the trajectory $\omega _{t}$, is given by 
\begin{equation}
P\big(0,(t,t\!+\!\overline{t}]|\omega _{t};\varrho \big)={\frac{p_{\ 0}^{t+%
\overline{t}}(j_{1},t_{1};\ldots ;j_{n},t_{n}|\varrho )}{%
p_{0}^{t}(j_{1},t_{1};\ldots ;j_{n},t_{n}|\varrho )}}  \label{eq:2.8}
\end{equation}%
$\big($cf.\ (\ref{eq:1.0b})$\big)$. By (\ref{eq:2.3}), (\ref{eq:2.5}), (\ref%
{eq:2.6}), we obtain immediately that the probability (\ref{eq:2.8}) can be
rewritten as 
\begin{equation}
P\big(0,(t,t\!+\!\overline{t}]|\omega _{t};\varrho \big)=\mathrm{Tr}\{%
\mathcal{S}(t\!+\!\overline{t},t)\varrho (t)\}=P_{\ t}^{t+\overline{t}}\big(%
0|\varrho (t)\big)\,,  \label{eq:2.9}
\end{equation}%
\begin{equation}
\varrho (t)={\frac{1}{C(t)}}\>\mathcal{S}(t,t_{n})\mathcal{J}_{j_{n}}(t_{n})%
\mathcal{S}(t_{n},t_{n-1})\cdots \mathcal{J}_{j_{1}}(t_{1})\mathcal{S}%
(t_{1},0)\varrho \,,  \label{eq:2.10}
\end{equation}%
where $C(t)$ is the normalization factor determined by $\mathrm{Tr}\{\varrho
(t)\}=1$ $\big($cf.\ (\ref{eq:1.0b}), (\ref{eq:1.0c}) and (\ref{eq:1.1})$%
\big)$. Similar results hold for the other EPDs conditioned upon some
trajectory up to time $t$. Therefore, all conditional probabilities can be
computed by (\ref{eq:2.5}) and (\ref{eq:2.6}) if one uses as initial state
the expression (\ref{eq:2.10}). Equation (\ref{eq:2.10}) gives the state of
the system at time $t$ conditioned upon the trajectory $\omega _{t}$ up to
time $t$ (the a posteriori state).

The interpretation of (\ref{eq:2.10}) is that, when no count is registered,
the system evolution is given by $\mathcal{S}(t,t_{0})$ and that the action
of the counter on the system, at the time $t$ in which a count of type $j$
is registered, is described by the map $\mathcal{J}_{j}(t)$. However, $%
\mathcal{S}(t,t_{0})$ and $\mathcal{J}_{j}(t)$ do not preserve normalization
and the normalization factor $C(t)$ is needed. This is due to the fact that
they are the probabilities (\ref{eq:2.5}) and (\ref{eq:2.6}) which have to
be correctly normalized and this is guaranteed by equations (\ref{eq:2.1})
and (\ref{eq:2.4}), connecting $\mathcal{J}_{j}(t)$ with $\mathcal{S}%
(t,t_{0})$. According to this interpretation of (\ref{eq:2.10}), the state
of the system in between two counts is 
\begin{equation}
\varrho (t)={\frac{\mathcal{S}(t,t_{r})\varrho (t_{r})}{\mathrm{Tr}\{%
\mathcal{S}(t,t_{r})\varrho (t_{r})\}}}\,,  \label{eq:2.11}
\end{equation}%
where $t_{r}$ is the time of the last count and $\varrho (t_{r})$ the state
just after this count. If we denote $\mathrm{Tr}\{R_{j}(t)\varrho (t)\}$ by $%
\langle R_{j}(t)\rangle _{t}$ and differentiate (\ref{eq:2.11}), we obtain 
\begin{equation}
{\frac{\mathrm{d}\varrho (t)}{\mathrm{d}t}}=\mathcal{L}_{0}(t)\varrho (t)-{%
\frac{1}{2}}\sum_{j=1}^{d}\big\{R_{j}(t)-\langle R_{j}(t)\rangle
_{t},\varrho (t)\big\}\,.  \label{eq:2.12}
\end{equation}%
Moreover, if at time $t_{r}$ we have a count of type $j$, the state of the
system soon after is 
\begin{equation}
\varrho (t_{r}\!+\!\mathrm{d}t)={\frac{\mathcal{J}_{j}(t_{r})\varrho (t_{r})%
}{\mathrm{Tr}\{\mathcal{J}_{j}(t_{r})\varrho (t_{r})\}}}={\frac{\mathcal{J}%
_{j}(t_{r})\varrho (t_{r})}{\langle R_{j}(t_{r})\rangle _{t_{r}}}}\,.
\label{eq:2.13}
\end{equation}

Now the typical trajectory $N_{j}(t)$ (number of counts of type $j$ up to
time $t$), $j=1,\ldots ,d$, of our stochastic process is a step function
such that $N_{j}(t)$ increases by 1 if soon after time $t$ there is a count
of type $j$, otherwise $N_{j}(t)$ is constant. Therefore, the It\^{o}
differential 
\begin{equation}
\mathrm{d}N_{j}(t)=N_{j}(t\!+\!\mathrm{d}t)-N_{j}(t)  \label{eq:2.14}
\end{equation}%
is equal to one if at time $t$ there is a count of type $j$ and to zero
otherwise. This gives $\big(\mathrm{d}N_{j}(t)\big)^{2}=\mathrm{d}N_{j}(t)$.
Moreover, the probability of more than one count in an interval $\mathrm{d}t$
vanishes more rapidly than $\mathrm{d}t$, i.e.\ between $\mathrm{d}N_{j}(t)$
and $\mathrm{d}N_{i}(t)$, $i\not=j$, at least one of the two must be zero.
Moreover, $\mathrm{d}N_{j}(t)\,\mathrm{d}t$ is of higher order than $\mathrm{%
d}t$ and has to be taken vanishing. Summarizing, we have the It\^{o} table 
\begin{equation}
\mathrm{d}N_{j}(t)\,\mathrm{d}N_{i}(t)=\delta _{ij}\,\mathrm{d}%
N_{j}(t)\,,\qquad \mathrm{d}N_{j}(t)\,\mathrm{d}t=0\,.  \label{eq:2.15}
\end{equation}%
By using these results we can rewrite (\ref{eq:2.12}) and (\ref{eq:2.13}) in
the form of a single stochastic differential equation in It\^{o} sense $\big(%
\,\mathrm{d}\varrho (t)=\varrho (t\!+\!\mathrm{d}t)-\varrho (t)\,\big)$: 
\begin{equation}
\mathrm{d}\varrho (t)=\mathcal{L}(t)\varrho (t)\,\mathrm{d}%
t+\sum_{j=1}^{d}\left( {\frac{\mathcal{J}_{j}(t)\varrho (t)}{\langle
R_{j}(t)\rangle _{t}}}-\varrho (t)\right) \Bigl(\mathrm{d}N_{j}(t)-\langle
R_{j}(t)\rangle _{t}\,\mathrm{d}t\Bigr)\,,  \label{eq:2.16}
\end{equation}%
\begin{equation}
\mathcal{L}(t)\varrho =\mathcal{L}_{0}(t)\varrho +\sum_{j=1}^{d}\Big(%
\mathcal{J}_{j}(t)\varrho -\frac{1}{2}\,\big\{R_{j}(t),\varrho \big\}\Big)\,,
\label{eq:2.17}
\end{equation}%
\begin{equation}
\langle R_{j}(t)\rangle _{t}=\mathrm{Tr}\big\{R_{j}(t)\varrho (t)\big\}=%
\mathrm{Tr}\big\{\mathcal{J}_{j}(t)\varrho (t)\big\}\,.  \label{eq:2.18}
\end{equation}%
Indeed, when all $\mathrm{d}N_{j}(t)$ vanish, (\ref{eq:2.16}) reduces to (%
\ref{eq:2.12}); when one of the $\mathrm{d}N_{j}(t)$ is equal to one all the
other terms in the r.h.s.\ of (\ref{eq:2.16}) are negligible and we obtain (%
\ref{eq:2.13}). Equation (\ref{eq:2.16}) was firstly obtained by quantum
stochastic calculus methods in \cite{bib:ref13,bib:ref19,bib:ref26,bib:ref21}%
.

Formula (\ref{eq:2.16}) is the equation for the a posteriori states in the
case of a counting measurement: it determines the state \emph{at time\/} $t$
depending on the (stochastic) trajectory \emph{up to time\/} $t$. Let us
stress that we know the solution of this equation: it is the state (\ref%
{eq:2.10}). In any case, it is very useful to have the differential
stochastic equation (\ref{eq:2.16}) as we shall see in the rest of this
section and in section \ref{sec:bmes3}.

Let $\langle \mathrm{d}N_{j}(t)\rangle (\omega _{t})$ be the mean number of
counts of type $j$ in the interval $(t,t\!+\!\mathrm{d}t]$ conditioned upon
the trajectory $\omega _{t}$ up to time $t$. Because probabilities of more
that one count in a small interval are negligible, we have 
\begin{equation}
\langle \mathrm{d}N_{j}(t)\rangle (\omega _{t})\simeq p_{\ t}^{t+\mathrm{d}t}%
\big(j,t|\varrho (t)\big)\,\mathrm{d}t\simeq \mathrm{Tr}\big\{\mathcal{J}%
_{j}(t)\varrho (t)\big\}\,\mathrm{d}t=\langle R_{j}(t)\rangle _{t}\,\mathrm{d%
}t\,.  \label{eq:2.19}
\end{equation}%
In other words the quantities $\langle R_{j}(t)\rangle _{t}\,\mathrm{d}t$
appearing in (\ref{eq:2.16}) are the \emph{a posteriori mean values\/} of $%
\mathrm{d}N_{j}(t)$. Moreover, the differentials 
\begin{equation}
\mathrm{d}M_{j}(t)=\mathrm{d}N_{j}(t)-\langle R_{j}(t)\rangle _{t}\,\mathrm{d%
}t\,,  \label{eq:2.19i}
\end{equation}%
appearing in (\ref{eq:2.16}), together with the initial condition $%
M_{j}(0)=0 $, define the a posteriori centered processes $M_{j}(t)$, called 
\emph{innovating martingales}.

Equation (\ref{eq:2.16}) is non--linear, but it is mathematically equivalent
to a linear one. Let us introduce an arbitrary stochastic real factor $c(t)$
and define the trace--class operator $\varphi(t) := c(t) \varrho(t)$. If we
know $\varphi(t)$ we can reobtain the state $\varrho(t)$ simply by
normalization. The factor $c(t)$ can be chosen in such a way that $%
\varphi(t) $ obeys a linear stochastic differential equation; moreover, this
choice is not unique. We shall do this in a very convenient way: a new
linear stochastic equation is obtained giving both the a posteriori state $%
\varrho(t)$ and the EPDs (\ref{eq:2.5}) and (\ref{eq:2.6}) (cf.\ \cite%
{bib:ref21}). Let $\varphi(t)$ be a trace--class operator depending on the
trajectory $\omega_t$ and defined by $\varphi(t) = \mathcal{S}%
(t,t_r)\varphi(t_r)$ if $t_r$ is the time of the last count and by $%
\varphi(t_r\!+\!\mathrm{d} t) = \tau \mathcal{J}_j(t_r)\varphi(t_r)$ if at
time $t_r$ there is a count of type $j$; $\tau$ is an arbitrary parameter
with dimensions of a time, which disappears from the physical quantities.
For initial condition we take $\varphi(0) = \varrho$.

By the definition of $\varphi (t)$, the quantity 
\begin{equation}
c(t)=\mathrm{Tr}\{\varphi (t)\}  \label{eq:2.19ii}
\end{equation}%
gives the EPDs (\ref{eq:2.5}) and (\ref{eq:2.6}): in the case of a
trajectory $\omega _{t}$ containing no jump we have 
\begin{equation}
c(t)=P_{0}^{t}(0|\varrho )  \label{eq:2.19iii}
\end{equation}%
and in the case of a trajectory with a jump of type $j_{1}$ at time $t_{1}$,
..., of type $j_{m}$ at time $t_{m}$ we have 
\begin{equation}
c(t)=\tau ^{m}\,p_{0}^{t}(j_{1},t_{1};\ldots ;j_{m},t_{m}|\varrho )\,.
\label{eq:2.19iv}
\end{equation}%
Moreover, by the definition of $\varphi (t)$, $c(t)$ and $\varrho (t)$ the a
posteriori state is 
\begin{equation}
\varrho (t)=\varphi (t)/c(t)\,.  \label{eq:2.19v}
\end{equation}

In the same way as for $\varrho(t)$, we can obtain the stochastic
differential equation for $\varphi(t)$, which turns out to be $\big($cf.\ 
\cite{bib:ref19}, equation (\ref{eq:??}20)$\big)$ 
\begin{equation}
\mathrm{d} \varphi(t) = \Bigl[ \mathcal{L}_0(t)\varphi(t) - \frac12\,
\sum_{j=1}^d \{R_j(t),\varphi(t)\} \Bigr]\mathrm{d} t + \sum_{j=1}^d \bigl[%
\tau\mathcal{J}_j(t)\varphi(t) - \varphi(t)\bigr] \mathrm{d} N_j(t)\,.
\label{eq:2.19vi}
\end{equation}
By using It\^o's calculus for counting processes it is possible to verify
that indeed (\ref{eq:2.19ii}), (\ref{eq:2.19v}) and (\ref{eq:2.19vi}) are
equivalent to (\ref{eq:2.16}). Equation (\ref{eq:2.19vi}) determines all the
probabilities via (\ref{eq:2.19ii})--(\ref{eq:2.19iv}) and the a posteriori
states via (\ref{eq:2.19ii}) and (\ref{eq:2.19v}). Equation (\ref{eq:2.19vi}%
) is linear, once a realization $N_j(t)$, $j=1,\cdots,d$, of the process is
given. However, let us note that the statistics of $N_j(t)$ depends in its
turn on the premeasurement state $\varrho$, as shown by (\ref{eq:2.5}), (\ref%
{eq:2.6}), (\ref{eq:2.19}). The possibility of finding a linear equation
mathematically equivalent to (\ref{eq:2.16}) means that $\varrho(t)$ is
linear in $\varrho$ up to a normalization factor, as suggested by (\ref%
{eq:1.1}).

Let us stress that in general equation (\ref{eq:2.16}) does not transform
pure states into pure states. This simply means that in the course of time
we loose information due to some dissipation mechanism, for instance the
system interacts also with some external bath or similar things. In any
case, the situation in which pure states are preserved is particularly
interesting. This is the case \cite{bib:XP26,bib:ref2,bib:ref13,bib:ref14}
when 
\begin{equation}
\mathcal{L}_{0}(t)\varrho =-\mathrm{i}[H(t),\varrho ]\,,\qquad \mathcal{J}%
_{j}(t)=Z_{j}(t)\varrho Z_{j}(t)^{\dagger }\,,  \label{eq:2.20.1}
\end{equation}%
where $Z_{j}(t)$ and $H(t)$ are operators on $\mathcal{H}$, $H(t)^{\dagger
}=H(t)$. Then, we have $R_{j}(t)=Z_{j}(t)^{\dagger }Z_{j}(t)$ and 
\begin{equation}
\mathcal{L}(t)\varrho =-\mathrm{i}[H(t),\varrho ]+\sum_{j=1}^{d}\left(
Z_{j}(t)\varrho Z_{j}(t)^{\dagger }-\frac{1}{2}\,\left\{ Z_{j}(t)^{\dagger
}Z_{j}(t),\varrho \right\} \right) \,.  \label{eq:2.20.2}
\end{equation}%
Then, (\ref{eq:2.16}) becomes 
\begin{eqnarray}
\mathrm{d}\varrho (t) &=&-\mathrm{i}[H(t),\varrho (t)]\mathrm{d}t-{\frac{1}{2%
}}\sum_{j=1}^{d}\left\{ Z_{j}(t)^{\dagger }Z_{j}(t)-\langle
Z_{j}(t)^{\dagger }Z_{j}(t)\rangle _{t},\varrho (t)\right\} \mathrm{d}t+{} 
\nonumber \\
&+&\sum_{j=1}^{d}\left( {\frac{Z_{j}(t)\varrho (t)Z_{j}(t)^{\dagger }}{%
\langle Z_{j}(t)^{\dagger }Z_{j}(t)\rangle _{t}}}-\varrho (t)\right) \mathrm{%
d}N_{j}(t)\,.  \label{eq:2.20.3}
\end{eqnarray}%
By using It\^{o} formula (\ref{eq:2.15}), one cane prove that $\varrho
(t\!+\!\mathrm{d}t)^{2}=\varrho (t\!+\!\mathrm{d}t)$, if $\varrho
(t)^{2}=\varrho (t)$; therefore, (\ref{eq:2.20.3}) transforms pure states
into pure states and it is equivalent to a stochastic differential equation
for a wave function. Indeed, let $\psi (t)\in \mathcal{H}$ satisfy the
\textquotedblleft a posteriori Schr\"{o}dinger equation" \cite%
{bib:ref19,bib:ref21} 
\begin{eqnarray}
\mathrm{d}\psi (t) &=&\Bigl[-\mathrm{i}H(t)-{\frac{1}{2}}\sum_{j=1}^{d}%
\left( Z_{j}(t)^{\dagger }Z_{j}(t)-\langle Z_{j}(t)^{\dagger
}Z_{j}(t)\rangle _{t}\right) \Bigr]\psi (t)\mathrm{d}t+{}  \nonumber \\
{} &+&\sum_{j=1}^{d}\left( {\frac{Z_{j}(t)}{\sqrt{\langle Z_{j}(t)^{\dagger
}Z_{j}(t)\rangle _{t}}}}-\mathbf{1}\right) \psi (t)\mathrm{d}N_{j}(t)\,.
\label{eq:2.20.4}
\end{eqnarray}%
with $\langle Z_{j}(t)^{\dagger }Z_{j}(t)\rangle _{t}=\langle \psi
(t)|Z_{j}(t)^{\dagger }Z_{j}(t)\,\psi (t)\rangle $; then, by (\ref{eq:2.15})
one obtains that $\varrho (t)=|\psi (t)\rangle \langle \psi (t)|$ satisfies (%
\ref{eq:2.20.3}).

It is interesting to note that in between two counts (when $\mathrm{d}
N_j(t) = 0$) (\ref{eq:2.20.4}) becomes a nonlinear Schr\"odinger equation of
the type studied, for instance, in \cite{bib:bmes40,bib:bhaml17}. However,
this equation has a quite different interpretation in the quoted references,
where the problem is to find evolution equations compatible with the Hilbert
space structure and preserving ``properties" in the sense of quantum logic.

Now we have the a posteriori states defined by (\ref{eq:2.16}) and a
probability measure on the trajectory space, which is implicitly defined by
the EPDs (\ref{eq:2.5}) and (\ref{eq:2.6}). Therefore, we can reconstruct
the instruments associated to our measurement by means of (\ref{eq:1.2}). As
in \cite{bib:ref19,bib:ref21}, we shall do this by using the notion of
characteristic operator, a concept introduced in \cite{bib:bhaml3}--\cite%
{bib:bel1}, and It\^o formula for counting processes.

Let $f$ be any function of the trajectories of our stochastic process and
let us denote by $\langle f\rangle _{\mathrm{st}}$ the mean value of $f$
with respect to the measure associated to the EPDs (\ref{eq:2.5}) and (\ref%
{eq:2.6}). The quantity 
\begin{equation}
\Phi _{t}[\vec{k}]=\bigg\langle\exp \biggl\{\mathrm{i}\sum_{j=1}^{d}%
\int_{0}^{t}k_{j}(s)\,\mathrm{d}N_{j}(s)\biggr\}\biggr\rangle_{\mathrm{st}}
\label{eq:2.20}
\end{equation}%
is called the \emph{characteristic functional\/} of the process. Here $\vec{k%
}(s)$ is a test function, i.e.\ $k_{j}(s)$ is a real compact support $%
C^{\infty }$-function on $(0,+\infty )$. $\Phi _{t}[\vec{k}]$ determines
uniquely the whole counting process up to time $t$: roughly speaking $\Phi
_{t}[\vec{k}]$ is the Fourier transform of the probability measure of the
process. More explicitly \cite{bib:bmes17}, we have 
\begin{eqnarray}
\lefteqn{\Phi _{t}[\vec{k}]=P_{0}^{t}(0|\varrho )+\sum_{m=1}^{\infty
}\sum_{\{j_{k}\}=1}^{\infty }\int_{0}^{t}\mathrm{d}t_{m}\int_{0}^{t_{m}}%
\mathrm{d}t_{m-1}\cdots }  \nonumber \\
&&\cdots \int_{0}^{t_{2}}\mathrm{d}t_{1}\,\exp \biggl\{\mathrm{i}%
\sum_{l=1}^{m}k_{j_{l}}(t_{l})\biggr\}\,p_{0}^{t}\big(%
j_{1},t_{1};j_{2},t_{2};\ldots ;j_{m},t_{m}|\varrho \big)\,.  \label{eq:2.21}
\end{eqnarray}

Let us set now 
\begin{equation}
V_{t}[\vec{k}]=\exp \biggl\{\mathrm{i}\sum_{j=1}^{d}\int_{0}^{t}k_{j}(s)\,%
\mathrm{d}N_{j}(s)\biggr\}\,.  \label{eq:2.22}
\end{equation}%
According to (\ref{eq:1.2}), we can write 
\begin{equation}
\langle V_{t}[\vec{k}]\varrho (t)\rangle _{\mathrm{st}}=\mathcal{G}_{t}[\vec{%
k}]\varrho \,,  \label{eq:2.23}
\end{equation}%
where $\varrho (t)$ is the a posteriori state at time $t$ and $\mathcal{G}%
_{t}[\vec{k}]$ is an operator on ${\mathcal{T}}({\mathcal{H}})$ which
represents the \textquotedblleft functional Fourier transform" of the
instrument $\mathcal{I}_{t}$ associated to our measurement up to time $t$.
The quantity $\mathcal{G}_{t}[\vec{k}]$ can be called \emph{characteristic
operator\/} and it is the operator analogue of the characteristic functional
of a stochastic process \cite{bib:bmes11}--\cite{bib:bel1}. By the
normalization of $\varrho (t)$ and (\ref{eq:2.20}) and (\ref{eq:2.23}), we
obtain 
\begin{equation}
\Phi _{t}[\vec{k}]=\mathrm{Tr}\big\{\mathcal{G}_{t}[\vec{k}]\varrho \big\}\,.
\label{eq:2.24}
\end{equation}

An equation for $\mathcal{G}_{t}[\vec{k}]$ can be found by differentiating (%
\ref{eq:2.23}). The differential of $\varrho (t)$ is given by (\ref{eq:2.16}%
), while the differential of $V_{t}[\vec{k}]$ is 
\begin{equation}
\mathrm{d}V_{t}[\vec{k}]=V_{t}[\vec{k}]\biggl[\sum_{j=1}^{d}\left( \mathrm{e}%
^{\mathrm{i}k_{j}(t)}-1\right) \,\mathrm{d}N_{j}(t)\biggr]\,.
\label{eq:2.25}
\end{equation}%
This formula can be easily obtained from (\ref{eq:2.22}) by expanding the
exponential and using (\ref{eq:2.15}). By using the formula 
\begin{equation}
\mathrm{d}\Big(V_{t}[\vec{k}]\varrho (t)\Big)=\left( \mathrm{d}V_{t}[\vec{k}%
]\right) \varrho (t)+V_{t}[\vec{k}]\bigl(\mathrm{d}\varrho (t)\bigr)+\left( 
\mathrm{d}V_{t}[\vec{k}]\right) \bigl(\mathrm{d}\varrho (t)\bigr)\,,
\label{eq:2.26}
\end{equation}%
where the \emph{It\^{o} correction\/} $\left( \mathrm{d}V_{t}[\vec{k}%
]\right) \left( \mathrm{d}\varrho (t)\right) $ has to be computed by means
of the It\^{o} table (\ref{eq:2.15}), we obtain 
\begin{eqnarray}
\mathrm{d}\Big(V_{t}[\vec{k}]\varrho (t)\Big) &=&V_{t}[\vec{k}]\bigg\{%
\mathcal{L}(t)\varrho (t)\,\mathrm{d}t+\sum_{j=1}^{d}\left( \mathrm{e}^{%
\mathrm{i}k_{j}(t)}-1\right) \mathcal{J}_{j}(t)\varrho (t)\,\mathrm{d}t+{} 
\nonumber \\
&&+\sum_{j=1}^{d}\left[ \mathrm{e}^{\mathrm{i}k_{j}(t)}\>{\frac{\mathcal{J}%
_{j}(t)\varrho (t)}{\langle R_{j}(t)\rangle _{t}}}-\varrho (t)\right] \Bigl(%
\mathrm{d}N_{j}(t)-\langle R_{j}(t)\rangle _{t}\,\mathrm{d}t\Bigr)\bigg\}\,.
\label{eq:2.27}
\end{eqnarray}

Now let us take the stochastic mean of (\ref{eq:2.27}). We compute this mean
in the following way. First we take the mean with respect to the probability
measure on the future (with respect to $t$) conditioned upon the given
trajectory. All the quantities in the r.h.s.\ of (\ref{eq:2.27}) depend only
on the past (they are \emph{adapted\/}), but the quantity $\mathrm{d}%
N_{j}(t) $, whose a posteriori mean value is just $\langle R_{j}(t)\rangle
_{t}\,\mathrm{d}t\ \big($equation (\ref{eq:2.19})$\big)$. Therefore, the
last term in (\ref{eq:2.27}) vanishes. Then, we take the mean value also on
the past and, by (\ref{eq:2.23}), we obtain 
\begin{equation}
{\frac{\mathrm{d}}{\mathrm{d}t}}\mathcal{G}_{t}[\vec{k}]=\mathcal{K}_{t}\big(%
\vec{k}(t)\big)\mathcal{G}_{t}[\vec{k}]\,,  \label{eq:2.28}
\end{equation}%
\begin{equation}
\mathcal{K}_{t}\big(\vec{k}(t)\big)=\mathcal{L}(t)+\sum_{j=1}^{d}\left( 
\mathrm{e}^{\mathrm{i}k_{j}(t)}-1\right) \mathcal{J}_{j}(t)\,.
\label{eq:2.29}
\end{equation}%
Together with the initial condition 
\begin{equation}
\mathcal{G}_{0}[\vec{k}]=\mathrm{Id}  \label{eq:2.30}
\end{equation}%
(which follows from the definition (\ref{eq:2.23})), equation (\ref{eq:2.28}%
) determines uniquely $\mathcal{G}_{t}[\vec{k}]$ and implicitly the
instruments on the trajectory space. This kind of equations has been
obtained for the first time in \cite{bib:bel1}.

If no selection is made according to the results of the measurement (let us
say: the results are not read), the state of the system at time $t$ will be $%
\big($cf.\ (\ref{eq:1.4})$\big)$ 
\begin{equation}
\sigma(t) = \langle\varrho(t)\rangle_{\mathrm{st}}\, ;  \label{eq:2.31}
\end{equation}
$\sigma(t)$ is the \emph{a priori state\/} for the case of the continuous
measurement described in this section. According to (\ref{eq:2.22}), (\ref%
{eq:2.23}), (\ref{eq:2.28}), (\ref{eq:2.29}), we have that the a priori
states satisfy the \emph{quantum master equation\/} 
\begin{equation}
{\frac{\mathrm{d} }{\mathrm{d} t}}\sigma(t) = \mathcal{L}(t)\sigma(t)\, ,
\label{eq:2.32}
\end{equation}
with the new Liouvillian (\ref{eq:2.17}): the unperturbed Liouvillian $%
\mathcal{L}_0(t)$ corrected by the measurement effect term $\sum_{j=1}^d\Big(%
\mathcal{J}_j(t)\varrho - \frac12\, \big\{R_j(t),\varrho\big\}\Big)$. The
fact that we have obtained a linear equation for the a priori states is due
to linearity and normalization of the instruments $\big($cf.\ (\ref{eq:1.4})$%
\big)$.

\section{An example of counting process:\newline
a two--level atom}

\typeout{An example of counting process: a two--level atom} \label{sec:bmes2}
\setcounter{equation}{0}

Let us consider an example of counting measurement on the simplest quantum
system: a two--state system, described in the Hilbert space $\mathcal{H} = 
\mathbf{C}^2$. We can think of a two--level atom, an unstable particle, a
spin... While the general case could be handled, for concreteness we treat a
two--level atom with pumping and damping. This section has to be considered
just as an illustration of the theory developed before.

The (time independent) unperturbed Liouvillian is given by 
\begin{equation}
\mathcal{L}_{0}\varrho =-\frac{\mathrm{i}}{2}\,\omega \lbrack \sigma
_{3},\varrho ]+\mathcal{J}_{0}\varrho -\frac{1}{2}\,\{R_{0},\varrho \}\,,
\label{eq:3.1}
\end{equation}%
\begin{equation}
\mathcal{J}_{0}\varrho =\lambda _{+}\,\sigma _{+}\varrho \sigma _{-}+\lambda
_{-}\,\sigma _{-}\varrho \sigma _{+}\,,  \label{eq:3.2}
\end{equation}%
\begin{equation}
R_{0}=\mathcal{J}_{0}^{\prime }\sigma _{0}=\frac{1}{2}\,\lambda _{+}(\sigma
_{0}-\sigma _{3})+\frac{1}{2}\,\lambda _{-}(\sigma _{0}+\sigma _{3})\,.
\label{eq:3.3}
\end{equation}%
Here $\omega >0$, $\lambda _{\pm }\geq 0$, $\sigma _{i}$, $i=1,2,3$, are the
Pauli matrices, $\sigma _{0}$ is the $2\times 2$ identity matrix and $\sigma
_{\pm }={\frac{1}{2}}(\sigma _{1}\pm \mathrm{i}\sigma _{2})$.

We consider a single counter $(d=1)$ and take 
\begin{equation}
\mathcal{J}_{1}\varrho =\lambda _{1}\,\sigma _{-}\varrho \sigma
_{+}\,,\qquad \lambda _{1}>0\,;  \label{eq:3.4}
\end{equation}%
the map $\mathcal{J}_{1}$ describes the emission of photons (or other types
of particles), which are then counted by some electronic device. In the
present case, the rate operator (\ref{eq:2.1}) is 
\begin{equation}
R_{1}=\mathcal{J}_{1}^{\prime }\sigma _{0}=\lambda _{1}\,\sigma _{+}\sigma
_{-}\equiv \frac{1}{2}\,\lambda _{1}(\sigma _{0}+\sigma _{3})  \label{eq:3.5}
\end{equation}%
and the generator $\mathcal{L}$ (\ref{eq:2.17}) of the full dynamics is 
\begin{equation}
\mathcal{L}\varrho =-\frac{\mathrm{i}}{2}\,\omega \lbrack \sigma
_{3},\varrho ]+\sum_{j=0}^{1}\left( \mathcal{J}_{j}\varrho -\frac{1}{2}%
\,\{R_{j},\varrho \}\right) \,.  \label{eq:3.6}
\end{equation}

We can interpret the terms with $\lambda_+$ as pumping, the terms with $%
\lambda_-$ as incoherent damping and the terms with $\lambda_1$ as
electromagnetic decay; $\Gamma = \lambda_1$ is the electromagnetic
transition rate. If $\lambda_+ = 0$, we can interpret the system as a Wigner
atom (or another unstable particle). In this case the electromagnetic
transition rate is $\Gamma = \lambda_- + \lambda_1$; $\lambda_- \not= 0$
means that not all the photons are collected by the photocounter; $%
\varepsilon = \lambda_1/(\lambda_- + \lambda_1)$ is the efficiency of the
counter \cite{bib:bmes17}.

In order to perform computations, it is convenient to represent selfadjoint
trace--class operators $\varphi$ as 
\begin{equation}
\varphi = \frac12\, \bigl(c\sigma_0 + \zeta \sigma_+ + \zeta^* \sigma_- +
\xi \sigma_3 \bigr)\,, \qquad c,\xi \in \mathrm{I\!R}\,, \quad \zeta \in%
\mathbf{C}\,.  \label{eq:3.7}
\end{equation}
The operator $\varphi$ is positive if $c \geq \left(\xi^2 +
\left|\zeta\right|^2 \right)^{1/2}$ and it is a density matrix if also $c =
1 $.

Let us consider (\ref{eq:2.19vi}) and represent $\varphi (t)$ in the form (%
\ref{eq:3.7}) with $c\rightarrow c(t)$, $\zeta \rightarrow \zeta (t)$, $\xi
\rightarrow \xi (t)$. The stochastic equation (\ref{eq:2.19vi}), choosing $%
\tau =\lambda _{1}^{\ -1}$, becomes 
\begin{equation}
\mathrm{d}c(t)+\frac{1}{2}\,\lambda _{1}[c(t)+\xi (t)]\mathrm{d}t=\frac{1}{2}%
\,[\xi (t)-c(t)]\mathrm{d}N(t)\,,  \label{eq:3.8}
\end{equation}%
\begin{equation}
\mathrm{d}\xi (t)+\left[ \left( 2\kappa -\frac{1}{2}\,\lambda _{1}\right)
\xi (t)+\left( \alpha +\frac{1}{2}\,\lambda _{1}\right) c(t)\right] \mathrm{d%
}t=-\frac{1}{2}\,[c(t)+3\xi (t)]\mathrm{d}N(t)\,,  \label{eq:3.9}
\end{equation}%
\begin{equation}
\mathrm{d}\zeta (t)+(\mathrm{i}\omega +\kappa )\zeta (t)\,\mathrm{d}t=-\zeta
(t)\,\mathrm{d}N(t)\,,  \label{eq:3.10}
\end{equation}%
where $\kappa ={\frac{1}{2}}(\lambda _{+}+\lambda _{-}+\lambda _{1})$, $%
\alpha =\lambda _{-}-\lambda _{+}$. It is convenient to rewrite (\ref{eq:3.8}%
) and (\ref{eq:3.9}) in terms of the stochastic parameters 
\begin{equation}
\pi _{0}(t)=\frac{1}{2}\,\bigl(c(t)-\xi (t)\bigr)\,,\qquad \pi _{1}(t)=\frac{%
1}{2}\,\bigl(c(t)+\xi (t)\bigr)\,;  \label{eq:3.11}
\end{equation}%
this gives 
\begin{eqnarray}
\mathrm{d}\pi _{0}(t)+[\mu _{\uparrow }\pi _{0}(t)-\kappa _{\downarrow }\pi
_{1}(t)]\mathrm{d}t &=&[\pi _{1}(t)-\pi _{0}(t)]\mathrm{d}N(t)\,,  \nonumber
\\
\mathrm{d}\pi _{1}(t)+[\mu _{\downarrow }\pi _{1}(t)-\kappa _{\uparrow }\pi
_{0}(t)]\mathrm{d}t &=&-\pi _{1}(t)\mathrm{d}N(t)\,,  \label{eq:3.12}
\end{eqnarray}%
where $\mu _{\uparrow }=\kappa _{\uparrow }=\lambda _{+}$, $\kappa
_{\downarrow }=\lambda _{-}$, $\mu _{\downarrow }=\lambda _{1}+\lambda _{-}$.

The solution of (\ref{eq:3.10}) is very simple: 
\begin{equation}
\zeta (t)=\left\{ 
\begin{array}{ll}
\mathrm{e}^{-(\mathrm{i}\omega +\kappa )t}\zeta (0), & \mbox{if $t\leq t_1$},
\\ 
0, & \mbox{if $t>t_1$},%
\end{array}%
\right.  \label{eq:3.13}
\end{equation}%
where $t_{1}$ is the instant of the first jump of $N(t)$. About (\ref%
{eq:3.12}), let us denote by $\pi _{j}(t|a,b)$ the solution of (\ref{eq:3.12}%
) with $\mathrm{d}N(t)=0$ and initial conditions $\pi _{0}(0)=a$, $\pi
_{1}(0)=b$. Then, the solution of the stochastic system (\ref{eq:3.12}) is 
\begin{equation}
\pi _{j}(t)=\left\{ 
\begin{array}{ll}
\pi _{j}\big(t|\pi _{0}(0),\pi _{1}(0)\big), & \mbox{if $t \leq t_1$}, \\ 
\pi _{j}\big(t-t_{r}|\pi _{1}(t_{r}),0\big), & 
\mbox{if $t_r < t \leq
t_{r+1}$, $r \geq 1$},%
\end{array}%
\right.  \label{eq:3.14}
\end{equation}%
where $t_{r}$ are the instants of the jumps of $N(t)$.

By (\ref{eq:2.19ii}), (\ref{eq:2.19v}), (\ref{eq:3.7}), the matrix elements
of the a posteriori state $\varrho (t)$ are given by 
\begin{eqnarray}
\langle 1|\varrho (t)|1\rangle &\equiv &\mathrm{Tr}\bigl\{\frac{1}{2}%
\,(\sigma _{0}+\sigma _{3})\varrho (t)\bigr\}=\pi _{1}(t)/c(t)\,,  \nonumber
\\
\langle 0|\varrho (t)|0\rangle &\equiv &\mathrm{Tr}\bigl\{\frac{1}{2}%
\,(\sigma _{0}-\sigma _{3})\varrho (t)\bigr\}=\pi _{0}(t)/c(t)\,,  \nonumber
\\
\langle 1|\varrho (t)|0\rangle &\equiv &\mathrm{Tr}\bigl\{\sigma _{-}\varrho
(t)\bigr\}=\zeta (t)/[2c(t)]\,,  \nonumber \\
\langle 0|\varrho (t)|1\rangle &\equiv &\mathrm{Tr}\bigl\{\sigma _{+}\varrho
(t)\bigr\}=\zeta (t)^{\ast }/[2c(t)]\,,  \label{eq:3.15}
\end{eqnarray}%
with $c(t)=\pi _{0}(t)+\pi _{1}(t)$. Equations (\ref{eq:3.13})--(\ref%
{eq:3.15}) shows that at a jump of $N(t)$ the system surely goes into the
ground state, because $\zeta =0$ and $\pi _{1}=0$, and that for $t>t_{1}$
the system is surely in a mixture of ground and excited states, because $%
\zeta =0$. The EPDs are implicitly given by $c(t)=\pi _{0}(t)+\pi _{1}(t)$, $%
\tau =\lambda _{1}^{-1}$, (\ref{eq:2.19iii}), (\ref{eq:2.19iv}), (\ref%
{eq:3.14}).

Just as an example let us discuss the case of the Wigner atom $(\lambda_+
=0) $. Equations (\ref{eq:3.14}) become 
\begin{equation}
\pi_1(t) = \left\{%
\begin{array}{ll}
\pi_0(0) + {\frac{\lambda_- }{2\kappa}} \left( 1 - \mathrm{e}^{-2\kappa
t}\right) \pi_1(0), & \mbox{if $t \leq t_1$}, \\ 
\exp[-2\kappa t_1]\, \pi_1(0), & \mbox{if $t_1 < t \leq t_2$}, \\ 
0, & \mbox{if $t > t_2$},%
\end{array}%
\right.  \label{eq:3.16}
\end{equation}
with $\kappa = {\frac{1}{2}} (\lambda_- + \lambda_1)$, $\pi_0(0) + \pi_1(0)
=1$. Equations (\ref{eq:3.15}) give $\varrho(t) = |0\rangle \langle0|$ for $%
t > t_1$: after the first registered emission the atom is in the ground
state. Finally the EPDs are 
\begin{equation}
P_0^t(0|\varrho) = \pi_0(0) + {\frac{1}{2\kappa}} \left(\lambda_- +
\lambda_1 \mathrm{e}^{-2\kappa t} \right) \pi_1(0)\,,  \label{eq:3.17}
\end{equation}
\begin{equation}
p_0^t(j_1,t_1|\varrho) = \lambda_1\, \exp[-2\kappa t_1]\, \pi_1(0)\,,
\label{eq:3.18}
\end{equation}
\begin{equation}
p_0^t\big(j_1,t_1; \ldots; j_m,t_m| \varrho\big) = 0\,, \qquad m \geq 2\,.
\label{eq:3.19}
\end{equation}
These equations say that there is at most a count, as it must be because
there is no pumping.

\section{Diffusion processes}

\typeout{Diffusion processes} \label{sec:bmes3} \setcounter{equation}{0}

In the classical case Gaussian diffusion processes can be obtained from
Poissonian counting ones by centering and scaling. Similarly, in the quantum
case we can obtain some kind of ``quantum diffusion measuring processes"
from the quantum counting processes of section \ref{sec:bmes1}.

Let us take the maps $\mathcal{J}_{j}(t)$, describing the action of the
counters, of the following form: 
\begin{equation}
\mathcal{J}_{j}(t)\varrho =\left[ Z_{j}(t)+\frac{1}{\varepsilon }\,f_{j}(t)%
\right] \varrho \left[ Z_{j}(t)^{\dagger }+\frac{1}{\varepsilon }%
\,f_{j}(t)^{\ast }\right] \,,  \label{eq:4.1}
\end{equation}%
where the $Z_{j}(t)$ are operators on ${\mathcal{H}}$, the $f_{j}$ are
complex functions and $\varepsilon >0$ is a parameter which we want to make
vanishing at the end. Moreover, instead of $\mathcal{L}_{0}(t)$ we take as
unperturbed Liouvillian the expression 
\begin{equation}
\mathcal{L}_{0}^{\varepsilon }(t)\varrho =\mathcal{L}_{0}(t)\varrho +{\frac{%
\mathrm{i}}{2\varepsilon }}\sum_{j=1}^{d}\left[ \mathrm{i}f_{j}(t)^{\ast
}Z_{j}(t)-\mathrm{i}f_{j}(t)Z_{j}(t)^{\dagger },\,\varrho \right] .
\label{eq:4.2}
\end{equation}%
Then, the generator $\mathcal{L}(t)$ of the a priori dynamics $\big($cf.\ (%
\ref{eq:2.17}) and (\ref{eq:2.32})$\big)$ becomes 
\begin{equation}
\mathcal{L}(t)\sigma =\mathcal{L}_{0}(t)\sigma +\sum_{j=1}^{d}\Bigl(%
Z_{j}(t)\sigma Z_{j}(t)^{\dagger }-\frac{1}{2}\,\left\{ Z_{j}(t)^{\dagger
}Z_{j}(t)\,,\,\sigma \right\} \Bigr)\,.  \label{eq:4.3}
\end{equation}%
The expression (\ref{eq:4.2}) has been assumed in order to have $\mathcal{L}%
(t)$ independent of the parameter $\varepsilon $. Physically, the structure (%
\ref{eq:4.1})--(\ref{eq:4.3}) is related to heterodyne detection \cite%
{bib:bmes42}.

Moreover, we make a linear transformation on the outputs: we call $%
Y_j^\varepsilon(t)$ the new observed processes, related to the old processes 
$N_j(t)$ by 
\begin{equation}
\mathrm{d} Y_j^\varepsilon(t)\!:\,= \varepsilon\, \mathrm{d} N_j(t) -
\frac1\varepsilon\, \left|f_j(t)\right|^2\, \mathrm{d} t\, ;  \label{eq:4.4}
\end{equation}
this means that we rescale the outputs and subtract a known deterministic
signal. Then, by (\ref{eq:4.4}) and (\ref{eq:2.15}) we obtain 
\begin{equation}
\mathrm{d} Y_j^\varepsilon(t)\, \mathrm{d} Y_i^\varepsilon(t) =
\varepsilon^2\delta_{ij}\, \mathrm{d} N_j(t) = \varepsilon\, \delta_{ij}\, 
\mathrm{d} Y_j^\varepsilon(t) + \delta_{ij} \left|f_j(t)\right|^2 \mathrm{d}
t\, .  \label{eq:4.5}
\end{equation}

In order to have the characteristic operator associated to this new
processes, we have to rescale the test function $\vec{k}(s)$, appearing in (%
\ref{eq:2.20}), (\ref{eq:2.24}), (\ref{eq:2.28})--(\ref{eq:2.30}), by
changing $k_{j}(t)$ into $\varepsilon \,k_{j}(t)$ and we have to shift the
mean values of $\varepsilon N_{j}(t)$ as in (\ref{eq:4.4}) by adding to $%
\mathcal{K}_{t}\big(\vec{k}(t)\big)$ the term $-{\frac{i}{\varepsilon }}%
\sum_{j}k_{j}(t)\left\vert f_{j}(t)\right\vert ^{2}$. The final result is
that the generator $\mathcal{K}_{t}\big(\vec{k}(t)\big)$ of the
characteristic operator $\mathcal{G}_{t}[\vec{k}]$ becomes 
\begin{eqnarray}
\mathcal{K}_{t}\big(\vec{k}(t)\big)\varrho &=&\mathcal{L}(t)\varrho
+\sum_{j=1}^{d}\biggl\{-\frac{1}{2}\,k_{j}(t)^{2}\left\vert
f_{j}(t)\right\vert ^{2}\varrho +\mathrm{i}k_{j}(t)\bigl[f_{j}(t)^{\ast
}Z_{j}(t)\varrho  \nonumber \\
&+&f_{j}(t)\varrho Z_{j}(t)^{\dagger }\bigr]+\left[ \mathrm{e}^{\mathrm{i}%
\varepsilon k_{j}(t)}-1\right] Z_{j}(t)\varrho Z_{j}(t)^{\dagger }  \nonumber
\\
&+&{\frac{1}{\varepsilon }}\left[ \mathrm{e}^{\mathrm{i}\varepsilon
k_{j}(t)}-1-\mathrm{i}\varepsilon k_{j}(t)\right] \bigl[f_{j}(t)^{\ast
}Z_{j}(t)\varrho +f_{j}(t)\varrho Z_{j}(t)^{\dagger }\bigr]  \nonumber \\
&+&{\frac{1}{\varepsilon ^{2}}}\left\vert f_{j}(t)\right\vert ^{2}\left[ 
\mathrm{e}^{\mathrm{i}\varepsilon k_{j}(t)}-1-\mathrm{i}\varepsilon k_{j}(t)+%
\frac{1}{2}\,\varepsilon ^{2}k_{j}(t)^{2}\right] \varrho \biggr\}\,.
\label{eq:4.6}
\end{eqnarray}

Also equation (\ref{eq:2.16}) for the a posteriori states can be expressed
in terms of the new processes $Y_{j}^{\varepsilon }(t)$. By (\ref{eq:2.1}), (%
\ref{eq:2.18}), (\ref{eq:4.1}), (\ref{eq:4.3}) and (\ref{eq:4.4}), we obtain 
\begin{eqnarray}
\mathrm{d}\varrho (t) &=&\mathcal{L}(t)\varrho (t)\,\mathrm{d}%
t+\sum_{j=1}^{d}\Bigl\{\varepsilon \,Z_{j}(t)\varrho Z_{j}(t)^{\dagger
}-\varepsilon \,\langle Z_{j}(t)^{\dagger }Z_{j}(t)\rangle _{t}\varrho (t)+ 
\nonumber \\
&+&f_{j}(t)^{\ast }\bigl[Z_{j}(t)-\langle Z_{j}(t)\rangle _{t}\bigr]\varrho
(t)+f_{j}(t)\varrho (t)\bigl[Z_{j}(t)^{\dagger }-\langle Z_{j}(t)^{\dagger
}\rangle _{t}\bigr]\Bigr\}  \nonumber \\
&\times &\Bigl[\varepsilon ^{2}\langle Z_{j}(t)^{\dagger }Z_{j}(t)\rangle
_{t}+\varepsilon f_{j}(t)^{\ast }\langle Z_{j}(t)\rangle _{t}+\varepsilon
f_{j}(t)\langle Z_{j}(t)^{\dagger }\rangle _{t}+\left\vert
f_{j}(t)\right\vert ^{2}\Bigr]^{-1}  \nonumber \\
&\times &\Bigl[\mathrm{d}Y_{j}^{\varepsilon }(t)-\varepsilon \langle
Z_{j}(t)^{\dagger }Z_{j}(t)\rangle _{t}\,\mathrm{d}t-f_{j}(t)^{\ast }\langle
Z_{j}(t)\rangle _{t}\,\mathrm{d}t  \nonumber \\
&&-f_{j}(t)\langle Z_{j}(t)^{\dagger }\rangle _{t}\,\mathrm{d}t\Bigr]\,,
\label{eq:4.7}
\end{eqnarray}%
where, for any operator $X$ on ${\mathcal{H}}$, $\langle X\rangle _{t}$ is
defined by 
\begin{equation}
\langle X\rangle _{t}=\mathrm{Tr}\{X\,\varrho (t)\}\,.  \label{eq:4.8}
\end{equation}%
Moreover, from (\ref{eq:4.4}), (\ref{eq:2.19}) and (\ref{eq:4.1}), we have
that the a posteriori mean values of $\mathrm{d}Y_{j}^{\varepsilon }(t)$ are
given by 
\begin{equation}
\langle \mathrm{d}Y_{j}^{\varepsilon }(t)\rangle (\omega _{t})=\bigl[%
f_{j}(t)\langle Z_{j}(t)^{\dagger }\rangle _{t}+f_{j}(t)^{\ast }\langle
Z_{j}(t)\rangle _{t}\bigr]\,\mathrm{d}t+\varepsilon \langle
Z_{j}(t)^{\dagger }Z_{j}(t)\rangle _{t}\,\mathrm{d}t\,.  \label{eq:4.9}
\end{equation}

We assume $\left\vert f_{j}(t)\right\vert \neq 0$, $\forall t$. From (\ref%
{eq:4.5})--(\ref{eq:4.7}) and (\ref{eq:4.9}), it is apparent that the limit $%
\varepsilon \downarrow 0$ exists. In this limit we obtain that the
characteristic operator is given by (\ref{eq:2.28}) and (\ref{eq:2.30}) with
generator 
\begin{eqnarray}
\lefteqn{\cK_{t}\big(\vec{k}(t)\big)\varrho =\cL(t)\varrho }  \nonumber \\
&&+\sum_{j=1}^{d}\Bigl\{-\frac{1}{2}\,k_{j}(t)^{2}\left\vert
f_{j}(t)\right\vert ^{2}\varrho +\mathrm{i}k_{j}(t)\bigl[f_{j}(t)^{\ast
}Z_{j}(t)\varrho +f_{j}(t)\,\varrho Z_{j}(t)^{\dagger }\bigr]\Bigr\}\,.
\label{eq:4.10}
\end{eqnarray}%
By setting $Y_{j}(t)=\lim_{\varepsilon \downarrow 0}Y_{j}^{\varepsilon }(t)$%
, the equation for the a posteriori states becomes 
\begin{eqnarray}
\mathrm{d}\varrho (t) &=&\mathcal{L}(t)\varrho (t)\,\mathrm{d}t  \nonumber \\
&+&\sum_{j=1}^{d}\Bigl\{f_{j}(t)^{\ast }\bigl[Z_{j}(t)-\langle
Z_{j}(t)\rangle _{t}\bigr]\varrho (t)+f_{j}(t)\varrho (t)\bigl[%
Z_{j}(t)^{\dagger }-\langle Z_{j}(t)^{\dagger }\rangle _{t}\bigr]\Bigr\} 
\nonumber \\
&\times &{\frac{1}{\left\vert f_{j}(t)\right\vert ^{2}}}\left[ \mathrm{d}%
Y_{j}(t)-f_{j}(t)^{\ast }\langle Z_{j}(t)\rangle _{t}\,\mathrm{d}%
t-f_{j}(t)\langle Z_{j}(t)^{\dagger }\rangle _{t}\,\mathrm{d}t\right] \,.
\label{eq:4.11}
\end{eqnarray}%
Moreover, the a posteriori mean value of $\mathrm{d}Y_{j}(t)$ becomes 
\begin{equation}
\langle \mathrm{d}Y_{j}(t)\rangle (\omega _{t})=2\mathrm{Re}\bigl[%
f_{j}(t)^{\ast }\langle Z_{j}(t)\rangle _{t}\bigr]\,\mathrm{d}t
\label{eq:4.12}
\end{equation}%
and the processes $M_{j}(t)$, defined by 
\begin{equation}
\mathrm{d}M_{j}(t)=\mathrm{d}Y_{j}(t)-2\mathrm{Re}\bigl[f_{j}(t)^{\ast
}\langle Z_{j}(t)\rangle _{t}\bigr]\,\mathrm{d}t\,,\qquad M_{j}(0)=0\,,
\label{eq:4.12i}
\end{equation}%
are again innovating martingales. Finally, the multiplication rule for the
differentials $\mathrm{d}Y_{j}(t)$ is the limit of (\ref{eq:4.5}) under $%
\varepsilon \downarrow 0$; by taking into account also the second of
equations (\ref{eq:2.15}), we have the It\^{o} table 
\begin{equation}
\mathrm{d}Y_{j}(t)\,\mathrm{d}Y_{i}(t)=\delta _{ji}\left\vert
f_{j}(t)\right\vert ^{2}\,\mathrm{d}t\,,\qquad \mathrm{d}Y_{j}(t)\,\mathrm{d}%
t=0\,.  \label{eq:4.13}
\end{equation}

By the procedure we have followed, it turns out that also the connection
between a posteriori states $\varrho (t)$ and characteristic operator $%
\mathcal{G}_{t}[\vec{k}]$ given by (\ref{eq:2.23}) continues to hold, but
now $\varrho (t)$ satisfies (\ref{eq:4.11}), $\mathcal{G}_{t}[\vec{k}]$
satisfies (\ref{eq:2.28}), (\ref{eq:2.30}) with generator given by (\ref%
{eq:4.10}) and $V_{t}[\vec{k}]$ is given by 
\begin{equation}
V_{t}[\vec{k}]=\exp \biggl[\mathrm{i}\sum_{j=1}^{d}\int_{0}^{t}k_{j}(s)\,%
\mathrm{d}Y_{j}(s)\biggr]\,.  \label{eq:4.14}
\end{equation}%
Alternatively, equation (\ref{eq:2.23}) can be proved by taking the
stochastic differential of both its sides, as done in the case of counting
processes.

By taking the mean value of (\ref{eq:4.12}) on the past, we obtain 
\begin{equation}
{\frac{\mathrm{d}}{\mathrm{d}t}}\langle Y_{j}(t)\rangle _{\mathrm{st}}=%
\mathrm{Tr}\Bigl\{\bigl[f_{j}(t)^{\ast
}\,Z_{j}(t)+f_{j}(t)\,Z_{j}(t)^{\dagger }\bigr]\sigma (t)\Bigr\}\,,
\label{eq:4.15}
\end{equation}%
where $\sigma (t)$ are the a priori states satisfying equation (\ref{eq:2.32}%
) with Liouvillian (\ref{eq:4.3}). The same result can be obtained by
functional differentiation of the characteristic functional $\mathrm{Tr}%
\left\{ \mathcal{G}_{T}[\vec{k}]\,\varrho \right\} $, $T>t$, with respect to 
$k_{j}(t)$ \cite{bib:bmes11}.

Equations (\ref{eq:4.14}) and (\ref{eq:4.15}) show us two things. First, our
continuous measurement gives the statistics of the generalized derivatives 
\cite{bib:bmes43} $y_j(t) = \dot Y_j(t)$ $\big($or of the increments $%
\mathrm{d} Y_j(t) \big)$ more than the statistics of the $Y_j(t)$
themselves. The same was true in the case of counting processes, but in that
case this difference was irrelevant, because we had the natural initial
condition $N_j(0) = 0$. Second, (\ref{eq:4.15}) can be interpreted by saying
that $y_j(t)$ is the output of a continuous measurement of the quantum
observables (selfadjoint operators) $f_j(t)^* Z_j(t) + f_j(t) Z_j(t)^\dagger$%
, which are in general noncommuting \cite{bib:bhaml3}--\cite%
{bib:bmes12,bib:XP31}--\cite{bib:ref13}.

Measuring processes defined by a characteristic operator with generator of
the type (\ref{eq:4.10}) were introduced in \cite{bib:bhaml3}--\cite%
{bib:bmes12} and equation (\ref{eq:4.11}) was obtained by quantum stochastic
calculus methods in \cite{bib:ref13}--\cite{bib:ref17,bib:ref26,bib:ref25}.
By linear transformations on the outputs, the most general diffusive case
can be reached; moreover, by taking prescription (\ref{eq:4.1}) only for a
subset of the $\mathcal{J}_j$, mixtures of diffusive and Poissonian
contributions can be obtained \cite{bib:ref26}. It\^o equations for the a
posteriori states in the \cite{bib:bel8,bib:bmes46,bib:bhaml18} purely
diffusive case have been considered also by Di\'osi \cite{bib:bel8}--\cite%
{bib:bhaml18}.

As in the case of counting processes there exists a (not unique) linear
stochastic equation mathematically equivalent to (\ref{eq:4.11}). For
instance, let $\varphi (t)$ be a trace class operator satisfying the
equation \cite{bib:ref26} 
\begin{equation}
\mathrm{d}\varphi (t)=\mathcal{L}(t)\varphi (t)\mathrm{d}t+\sum_{j=1}^{d}%
\left[ {\frac{1}{f_{j}(t)}}Z_{j}(t)\varphi (t)+{\frac{1}{f_{j}(t)^{\ast }}}%
\varphi (t)Z_{j}(t)^{\dagger }\right] \mathrm{d}Y_{j}(t)  \label{eq:4.15i}
\end{equation}%
and set $c(t):=\mathrm{Tr}\{\varphi (t)\}$. Then, by It\^{o}'s calculus one
shows that $\varrho (t)=\varphi (t)/c(t)$ satisfies (\ref{eq:4.11}). To the
linear equation (\ref{eq:4.15i}) the same comments apply as to (\ref%
{eq:2.19vi}).

In the case of an unperturbed Liouvillian of a purely Hamiltonian form, 
\begin{equation}
\mathcal{L}_{0}(t)\,\varrho =-\mathrm{i}[H(t),\varrho ]\,,  \label{eq:4.16}
\end{equation}%
equation (\ref{eq:4.11}) transforms pure states into pure ones; for proving
this it is sufficient to show that $\varrho (t+\mathrm{d}t)^{2}=\varrho (t+%
\mathrm{d}t)$ if $\varrho (t)^{2}=\varrho (t)$. In this case, which we can
call of complete measurement, (\ref{eq:4.11}) is equivalent to a stochastic
differential equation for a wave function, as in the case of counting
measurements. Indeed, let $\psi (t)\in \mathcal{H}$ satisfy the
\textquotedblleft a posteriori Schr\"{o}dinger equation" \cite%
{bib:ref14,bib:ref25} 
\begin{eqnarray}
\mathrm{d}\psi (t) &=&\!-\Bigl\{\mathrm{i}H(t)\!+{\frac{1}{2}}\sum_{j=1}^{d}%
\bigl[Z_{j}(t)^{\dagger }Z_{j}(t)-2\langle Z_{j}(t)^{\dagger }\rangle
_{t}Z_{j}(t)+\left\vert \langle Z_{j}(t)\rangle _{t}\right\vert ^{2}\bigr]%
\Bigr\}\psi (t)\mathrm{d}t  \nonumber \\
&+&\!\sum_{j=1}^{d}{\frac{1}{f_{j}(t)}}\bigl[Z_{j}(t)-\langle
Z_{j}(t)\rangle _{t}\bigr]\psi (t)\bigr[\mathrm{d}Y_{j}(t)-f_{j}(t)^{\ast
}\langle Z_{j}(t)\rangle _{t}\mathrm{d}t  \nonumber \\
&&-f_{j}(t)\langle Z_{j}(t)^{\dagger }\rangle _{t}\mathrm{d}t\bigr]\,,
\label{eq:4.17}
\end{eqnarray}%
with $\langle Z_{j}(t)\rangle _{t}=\langle \psi (t)|Z_{j}(t)\,\psi
(t)\rangle $; then, by It\^{o}'s calculus, one obtains that $\varrho
(t)\equiv |\psi (t)\rangle \langle \psi (t)|$ satisfies (\ref{eq:4.11}).

It is interesting to note that stochastic equations of the type of (\ref%
{eq:4.11}) and (\ref{eq:4.17}), with $f_j(t) = 1$, have been appeared in the
literature also in connection with \emph{dynamical theories of
wave--function reduction\/} \cite{bib:bmes48}--\cite{bib:bmes52}. The idea
is that the wave--function reduction associated to a measurement is some
kind of stochastic process and an equation of the type of (\ref{eq:4.17}) is
postulated. Apart from the different interpretations, another important
difference is that in the dynamical reduction theories the noise comes from
outside, while for us it is determined by the system itself.

Sometimes it is useful to have at disposal a complexified version of
diffusion processes. Let us consider the case of an even $d$. By redefining $%
d$ and the index $j$, the sum appearing in (\ref{eq:4.10}) and (\ref{eq:4.11}%
) can be reorganized as a double sum over $\lambda $, $\lambda =1,2$, and $j$%
, $j=1,\ldots ,d$. Then, we take $f_{1j}(t)=1$, $f_{2j}(t)=\mathrm{i}$, $%
Z_{1j}(t)=Z_{2j}(t)\equiv Z_{j}(t)$ and set $\kappa _{j}(t)=k_{1j}(t)+%
\mathrm{i}k_{2j}(t)$, $W_{j}(t)={\frac{1}{2}}\left( Y_{1j}(t)+\mathrm{i}%
Y_{2j}(t)\right) $. Then, (\ref{eq:4.3}), (\ref{eq:4.10})--(\ref{eq:4.12}), (%
\ref{eq:4.14}) become 
\begin{equation}
\mathcal{L}(t)\sigma =\mathcal{L}_{0}(t)\sigma +2\sum_{j=1}^{d}\Bigl(%
Z_{j}(t)\sigma Z_{j}(t)^{\dagger }-\frac{1}{2}\,\left\{ Z_{j}(t)^{\dagger
}Z_{j}(t)\,,\,\sigma \right\} \Bigr)\,,  \label{eq:4.30}
\end{equation}%
\begin{equation}
\mathcal{K}_{t}\big(\vec{\kappa}(t)\big)\varrho =\mathcal{L}(t)\varrho
+\sum_{j=1}^{d}\Bigl\{-\frac{1}{2}\,\left\vert \kappa _{j}(t)\right\vert
^{2}\varrho +\mathrm{i}\bigl[\kappa _{j}(t)^{\ast }Z_{j}(t)\varrho +\kappa
_{j}(t)\,\varrho Z_{j}(t)^{\dagger }\bigr]\Bigr\}\,,  \label{eq:4.31}
\end{equation}%
\begin{eqnarray}
\mathrm{d}\varrho (t) &=&\mathcal{L}(t)\varrho (t)\,\mathrm{d}%
t+2\sum_{j=1}^{d}\Bigl\{\bigl[Z_{j}(t)-\langle Z_{j}(t)\rangle _{t}\bigr]%
\varrho (t)\bigl[\mathrm{d}W_{j}(t)^{\ast }-\langle Z_{j}(t)^{\dagger
}\rangle _{t}\mathrm{d}t\bigr]  \nonumber \\
&+&\bigl[\mathrm{d}W_{j}(t)-\langle Z_{j}(t)\rangle _{t}\mathrm{d}t\bigr]%
\varrho (t)\bigl[Z_{j}(t)^{\dagger }-\langle Z_{j}(t)^{\dagger }\rangle _{t}%
\bigr]\Bigr\}\,,  \label{eq:4.32}
\end{eqnarray}%
\begin{equation}
\mathrm{d}W_{j}(t)\,\mathrm{d}W_{i}(t)=0\,,\quad \mathrm{d}W_{j}(t)^{\ast }\,%
\mathrm{d}W_{i}(t)=\frac{1}{2}\,\delta _{ji}\,\mathrm{d}t\,,\quad \mathrm{d}%
W_{j}(t)\,\mathrm{d}t=0\,,  \label{eq:4.33}
\end{equation}%
\begin{equation}
\langle \mathrm{d}W_{j}(t)\rangle (\omega _{t})=\langle Z_{j}(t)\rangle _{t}%
\mathrm{d}t\,.  \label{eq:4.34}
\end{equation}%
By taking the mean value of (\ref{eq:4.34}) on the past, we obtain 
\begin{equation}
{\frac{\mathrm{d}}{\mathrm{d}t}}\langle W_{j}(t)\rangle _{\mathrm{st}}=%
\mathrm{Tr}\bigl\{Z_{j}(t)\,\sigma (t)\bigr\}\,,  \label{eq:4.36}
\end{equation}%
which allows to interpret the equations above as describing a continuous
measurement of the noncommuting, nonselfadjoint operators $Z_{j}(t)$.
Filtering equation (\ref{eq:4.32}) for linear systems (quantum oscillators)
was introduced in \cite{bib:ref2,bib:XP31}.

\section{An example of diffusion process}

\typeout{An example of diffusion process} \label{sec:bmes4} %
\setcounter{equation}{0}

Let us close the paper with a simple example of the theory developed in
Section \ref{sec:bmes3}, in the complexified version (\ref{eq:4.30})--(\ref%
{eq:4.36}). A real--valued Gaussian example for an observed particle in a
quadratic potential can be found in \cite{bib:ref13,bib:ref18}. We consider
a single--mode field in a cavity and with a source, 
\begin{equation}
H(t)=\omega a^{\dagger }a+g(t)a^{\dagger }+g(t)^{\ast }a\,,\qquad \omega
>0\,,  \label{eq:5.1}
\end{equation}%
interacting with a thermal bath, 
\begin{equation}
\mathcal{L}_{0}(t)\varrho =-\mathrm{i}[H(t),\varrho ]+\lambda _{\downarrow }%
\bigl([a\varrho ,a^{\dagger }]+[a,\varrho a^{\dagger }]\bigr)+\lambda
_{\uparrow }\bigl([a^{\dagger }\varrho ,a]+[a^{\dagger },\varrho a]\bigr)\,,
\label{eq:5.2}
\end{equation}%
$\lambda _{\downarrow },\lambda _{\uparrow }\geq 0$, and subjected to the
measurement of a single complex observable ($d=1$) proportional to the
annihilation operator, 
\begin{equation}
Z=\eta a\,,\qquad \eta \in \mathbf{C}\,.  \label{eq:5.3}
\end{equation}%
The fact that $Z$ is proportional to $a$ means that we are considering a
passive, purely absorbing detector.

By scaling the output in such a way that we have exactly a measurement of $a$
$\big(\mathrm{d}W(t)/\eta \rightarrow \mathrm{d}W(t)$, $\eta ^{\ast }\kappa
(t)\rightarrow \kappa (t)\big)$, equations (\ref{eq:4.30})--(\ref{eq:4.36})
become 
\begin{equation}
\mathcal{L}(t)\sigma =\mathcal{L}_{0}(t)\sigma +\left\vert \eta \right\vert
^{2}\left( [a\sigma ,a^{\dagger }]+[a,\sigma a^{\dagger }]\right) \,,
\label{eq:5.4}
\end{equation}%
\begin{equation}
\mathcal{K}_{t}\big(\kappa (t)^{\ast },\kappa (t)\big)\varrho =\mathcal{L}%
(t)\varrho -\frac{1}{2}\,\left\vert \kappa (t)/\eta \right\vert ^{2}\varrho +%
\mathrm{i}\left[ \kappa (t)^{\ast }a\varrho +\kappa (t)\varrho a^{\dagger }%
\right] \,,  \label{eq:5.5}
\end{equation}%
\begin{eqnarray}
\mathrm{d}\varrho (t) &=&\mathcal{L}(t)\varrho (t)\mathrm{d}t+2\left\vert
\eta \right\vert ^{2}\Bigl\{\left[ a-\langle a\rangle _{t}\right] \varrho (t)%
\left[ \mathrm{d}W(t)^{\ast }-\langle a^{\dagger }\rangle _{t}\mathrm{d}t%
\right]  \nonumber \\
&&+\left[ \mathrm{d}W(t)-\langle a\rangle _{t}\mathrm{d}t\right] \varrho (t)%
\left[ a^{\dagger }-\langle a^{\dagger }\rangle _{t}\right] \Bigr\}\,,
\label{eq:5.6}
\end{eqnarray}%
\begin{equation}
\bigl(\mathrm{d}W(t)\bigr)^{2}=0\,,\qquad \left\vert \mathrm{d}%
W(t)\right\vert ^{2}=\mathrm{d}t\big/\bigl(2\left\vert \eta \right\vert ^{2}%
\bigr)\,,\qquad \mathrm{d}W(t)\,\mathrm{d}t=0\,,  \label{eq:5.7}
\end{equation}%
\begin{equation}
\mathrm{d}\langle W(t)\rangle (\omega _{t})=\langle a\rangle _{t}\mathrm{d}%
t\,,\qquad {\frac{\mathrm{d}}{\mathrm{d}t}}\langle W(t)\rangle _{\mathrm{st}%
}=\mathrm{Tr}\{a\,\sigma (t)\}\,.  \label{eq:5.8}
\end{equation}

Equations (\ref{eq:2.28}), with generator (\ref{eq:5.5}), and (\ref{eq:5.6})
can be solved by antinormal ordering expansion of trace--class operators.
Let us define on $\mathcal{T}(\mathcal{H})$ a \textquotedblleft tilde"
operation by $\varphi \in \mathcal{T}(\mathcal{H})\longrightarrow \tilde{%
\varphi}(\xi ^{\ast },\xi )$, 
\begin{equation}
\tilde{\varphi}(\xi ^{\ast },\xi )=\mathrm{Tr}\left\{ \mathrm{e}^{-\mathrm{i}%
\xi ^{\ast }a}\,\varphi \,\mathrm{e}^{-\mathrm{i}\xi a^{\dagger }}\right\}
\,,  \label{eq:5.9}
\end{equation}%
which can be inverted by 
\begin{equation}
\varphi ={\frac{1}{\pi }}\int \mathrm{d}_{2}\xi \,\mathrm{e}^{\mathrm{i}\xi
^{\ast }a}\,\mathrm{e}^{\mathrm{i}\xi a^{\dagger }}\,\tilde{\varphi}(\xi
^{\ast },\xi )\,.  \label{eq:5.10}
\end{equation}

Let us set 
\begin{equation}
\varphi (t)\equiv \varphi (\kappa ^{\ast },\kappa ;t)=\mathcal{G}_{t}[\kappa
^{\ast },\kappa ]\,\varrho \,;  \label{eq:5.11}
\end{equation}%
then (\ref{eq:2.28}) and (\ref{eq:5.5}) give, by standard computations, 
\begin{eqnarray}
{\frac{\partial }{\partial t}}\tilde{\varphi}(\xi ^{\ast },\xi ;t) &=&\Bigl[%
-\left( \mathrm{i}\omega +\frac{1}{2}\,\Gamma \right) \xi ^{\ast }\partial
^{\ast }+\left( \mathrm{i}\omega -\frac{1}{2}\,\Gamma \right) \xi \partial
-\kappa (t)^{\ast }\partial ^{\ast }-\kappa (t)\partial  \nonumber \\
&-&2\lambda _{\uparrow }\left\vert \xi \right\vert ^{2}-g(t)\xi ^{\ast
}+g(t)^{\ast }\xi -\frac{1}{2}\,\left\vert \kappa (t)/\eta \right\vert ^{2}%
\Bigr]\tilde{\varphi}(\xi ^{\ast },\xi ;t)\,,  \label{eq:5.12}
\end{eqnarray}%
where $\partial =\partial /\partial \xi $, $\partial ^{\ast }=\partial
/\partial \xi ^{\ast }$ and 
\begin{equation}
\Gamma =2\bigl(\left\vert \eta \right\vert ^{2}+\lambda _{\downarrow
}-\lambda _{\uparrow }\bigr)\,.  \label{eq:5.13}
\end{equation}%
We suppose $\Gamma $ be strictly positive.

If the initial condition is \textquotedblleft Gaussian", 
\begin{equation}
\tilde{\varrho}(\xi ^{\ast },\xi )=\exp \left[ -\mathrm{i}(\xi ^{\ast
}\alpha _{0}+\xi \alpha _{0}^{\ast })-\frac{1}{2}\,\left( \xi ^{\ast 2}\mu
_{0}+\xi ^{2}\mu _{0}^{\ast }\right) -\left\vert \xi \right\vert ^{2}\nu _{0}%
\right] \,,  \label{eq:5.14}
\end{equation}%
then $\tilde{\varphi}$ maintains this structure at any time. Indeed, by
inserting 
\begin{equation}
\tilde{\varphi}(\xi ^{\ast },\xi ;t)=\exp \Bigl\{-\mathrm{i}\left[ \xi
^{\ast }b(t)+\xi c(t)^{\ast }\right] -\frac{1}{2}\,\left[ \xi ^{\ast
2}d(t)+\xi ^{2}d(t)^{\ast }\right] -\left\vert \xi \right\vert ^{2}f(t)-h(t)%
\Bigr\}  \label{eq:5.15}
\end{equation}%
into (\ref{eq:5.12}), we obtain the differential equations for the
coefficients ($f$ is real): 
\begin{eqnarray}
\dot{b}(t) &=&-\left( \mathrm{i}\omega +\frac{1}{2}\,\Gamma \right) b(t)+%
\mathrm{i}\kappa (t)^{\ast }d(t)+\mathrm{i}\kappa (t)f(t)-\mathrm{i}g(t)\,, 
\nonumber \\
\dot{c}(t) &=&-\left( \mathrm{i}\omega +\frac{1}{2}\,\Gamma \right) c(t)-%
\mathrm{i}\kappa (t)^{\ast }d(t)-\mathrm{i}\kappa (t)f(t)-\mathrm{i}g(t)\,, 
\nonumber \\
\dot{d}(t) &=&-(2\mathrm{i}\omega +\Gamma )d(t)\,,  \nonumber \\
\dot{f}(t) &=&-\Gamma f(t)+2\lambda _{\uparrow }\,,  \nonumber \\
\dot{h}(t) &=&-\mathrm{i}\kappa (t)^{\ast }b(t)-\mathrm{i}\kappa
(t)c(t)^{\ast }+\frac{1}{2}\,\left\vert \kappa (t)/\eta \right\vert ^{2}\,.
\label{eq:5.16}
\end{eqnarray}%
The solution of these equations can be easily written down.

The characteristic functional of our generalized process \cite{bib:bmes43} $%
\dot{W}(t)$ is given by $\big($see (\ref{eq:2.24}) and (\ref{eq:2.30})$\big)$
\begin{equation}
\Phi _{t}[\kappa ^{\ast },\kappa ]=\mathrm{Tr}\bigl\{\varphi (\kappa ^{\ast
},\kappa ;t)\bigr\}=\tilde{\varphi}(0,0;t)=\exp [h(t)]  \label{eq:5.17}
\end{equation}%
with 
\begin{eqnarray}
h(t) &=&-\mathrm{i}\int_{0}^{t}\mathrm{d}s\left[ \kappa (s)^{\ast }\alpha
(s)+\kappa (s)\alpha (s)^{\ast }\right] +\int_{0}^{t}\mathrm{d}s\,\mathrm{d}%
s^{\prime }\,\Bigl[\kappa (s)^{\ast }\kappa (s^{\prime })\Delta
_{1}(s,s^{\prime })  \nonumber \\
&+&\frac{1}{2}\,\kappa (s)\kappa (s^{\prime })\Delta _{2}(s,s^{\prime
})^{\ast }+\frac{1}{2}\,\kappa (s)^{\ast }\kappa (s^{\prime })^{\ast }\Delta
_{2}(s,s^{\prime })\Bigr]\,,  \label{eq:5.18}
\end{eqnarray}%
\begin{equation}
\alpha (t)=\mathrm{e}^{-(\mathrm{i}\omega +{\frac{1}{2}}\Gamma )t}\alpha
_{0}-\mathrm{i}\int_{0}^{t}\mathrm{d}s\,g(s)\,\mathrm{e}^{-(\mathrm{i}\omega
+\Gamma /2)(t-s)}\,,  \label{eq:5.19}
\end{equation}%
\begin{eqnarray}
\Delta _{1}(s,s^{\prime }) &=&{\frac{1}{2\left\vert \eta \right\vert ^{2}}}%
\delta (s-s^{\prime })+\vartheta (s-s^{\prime })\,\mathrm{e}^{-(\mathrm{i}%
\omega +\Gamma /2)(s-s^{\prime })}\,C(s^{\prime })  \nonumber \\
&&+\vartheta (s^{\prime }-s)\,\mathrm{e}^{(\mathrm{i}\omega -\Gamma
/2)(s^{\prime }-s)}\,C(s)\,,  \label{eq:5.20}
\end{eqnarray}%
\begin{equation}
C(s)={\frac{2\lambda _{\uparrow }}{\Gamma }}+\left( \nu _{0}-{\frac{2\lambda
_{\uparrow }}{\Gamma }}\right) \mathrm{e}^{-\Gamma s}\,,  \label{eq:5.20i}
\end{equation}%
\begin{equation}
\Delta _{2}(s,s^{\prime })=\mathrm{e}^{-(\mathrm{i}\omega +\Gamma
/2)(s+s^{\prime })}\,\mu _{0}\,,  \label{eq:5.21}
\end{equation}%
where $\vartheta $ is the usual step function. $\Phi _{T}[\kappa ^{\ast
},\kappa ]$ is the characteristic functional of a Gaussian complex process
with covariance (\ref{eq:5.20}), (\ref{eq:5.21}) and (a priori) mean values 
\begin{equation}
{\frac{\mathrm{d}}{\mathrm{d}t}}\langle W(t)\rangle _{\mathrm{st}}=\mathrm{i}%
{\frac{\delta }{\delta \kappa (t)}}\Phi _{t}[\kappa ^{\ast },\kappa ]\Big|%
_{\kappa =\kappa ^{\ast }=0}=\mathrm{Tr}\bigl\{a\sigma (t)\bigr\}=\alpha
(t)\,.  \label{eq:5.22}
\end{equation}

The a priori states are given by $\sigma (t)=\mathcal{G}_{t}[0]\,\varrho $
or $\tilde{\sigma}(\xi ^{\ast },\xi ;t)=\newline
\tilde{\varphi}(\xi ^{\ast },\xi ;t)\Big|_{\kappa =\kappa ^{\ast }=0}$. By (%
\ref{eq:5.15}) and (\ref{eq:5.16}) we obtain 
\begin{equation}
\tilde{\sigma}(\xi ^{\ast },\xi ;t)=\exp \left\{ -\mathrm{i}\left[ \xi
^{\ast }\alpha (t)+c.c.\right] -\frac{1}{2}\,\left[ \xi ^{\ast 2}\mathrm{e}%
^{-(2\mathrm{i}\omega +\Gamma )t}\,\mu _{0}+c.c.\right] -\left\vert \xi
\right\vert ^{2}C(t)\right\} \,.  \label{eq:5.22i}
\end{equation}%
This gives 
\begin{equation}
\mathrm{Tr}\left\{ a^{\dagger }a\,\sigma (t)\right\} =C(t)\,,\quad \mathrm{Tr%
}\left\{ a^{2}\,\sigma (t)\right\} =\exp \left[ -2\left( \mathrm{i}\omega +%
\frac{1}{2}\,\Gamma \right) t\right] \mu _{0}\,.  \label{eq:5.22ii}
\end{equation}%
Note the links between the covariance (\ref{eq:5.22ii}) of the a priori
states $\sigma (t)$ and the covariance (\ref{eq:5.20}), (\ref{eq:5.21}) of
the process $\dot{W}(t)$.

By the \textquotedblleft tilde" transformation (\ref{eq:5.9}), we can solve
also the equation for the a posteriori states (\ref{eq:5.6}). From (\ref%
{eq:5.6}), (\ref{eq:5.9}), (\ref{eq:5.10}) we obtain 
\begin{eqnarray}
\mathrm{d}\tilde{\varrho}(\xi ^{\ast },\xi ;t) &=&\Bigl[-\left( \mathrm{i}%
\omega +\frac{1}{2}\,\Gamma \right) \xi ^{\ast }\partial ^{\ast }+\left( 
\mathrm{i}\omega -\frac{1}{2}\,\Gamma \right) \xi \partial -2\lambda
_{\uparrow }\left\vert \xi \right\vert ^{2}-g(t)\xi ^{\ast }  \nonumber \\
&+&g(t)^{\ast }\xi \Bigr]\tilde{\varrho}(\xi ^{\ast },\xi ;t)\,\mathrm{d}%
t+2\left\vert \eta \right\vert ^{2}\Bigl\{\left[ \mathrm{d}W(t)^{\ast
}-\langle a^{\dagger }\rangle _{t}\mathrm{d}t\right] \left[ \mathrm{i}%
\partial ^{\ast }-\langle a\rangle _{t}\right]  \nonumber \\
&+&\left[ \mathrm{d}W(t)-\langle a\rangle _{t}\mathrm{d}t\right] \left[ 
\mathrm{i}\partial -\langle a^{\dagger }\rangle _{t}\right] \Bigr\}\tilde{%
\varrho}(\xi ^{\ast },\xi ;t)\,.  \label{eq:5.23}
\end{eqnarray}%
This equation can be rewritten in terms of the stochastic function 
\begin{equation}
l(\xi ^{\ast },\xi ;t)\!:\,=-\mathrm{ln}\,\tilde{\varrho}(\xi ^{\ast },\xi
;t)\,.  \label{eq:5.24}
\end{equation}%
By using It\^{o}'s formula $\mathrm{d}\tilde{\varrho}/\tilde{\varrho}=-%
\mathrm{d}l+{\frac{1}{2}}(\mathrm{d}l)^{2}$, which in turn implies $(\mathrm{%
d}\tilde{\varrho}/\tilde{\varrho})^{2}=(\mathrm{d}l)^{2}$, and It\^{o}'s
table (\ref{eq:5.7}), we obtain 
\begin{eqnarray}
\mathrm{d}l &=&\Bigl[-2\left\vert \eta \right\vert ^{2}\partial ^{\ast
}l\,\partial l-\left( \mathrm{i}\omega +\frac{1}{2}\,\Gamma \right) \xi
^{\ast }\partial ^{\ast }l+\left( \mathrm{i}\omega -\frac{1}{2}\,\Gamma
\right) \xi \partial l  \nonumber \\
&+&2\lambda _{\uparrow }\left\vert \xi \right\vert ^{2}+g(t)\xi ^{\ast
}-g(t)^{\ast }\xi -2\left\vert \eta \right\vert ^{2}\left\vert \langle
a\rangle _{t}\right\vert ^{2}\Bigr]\mathrm{d}t  \nonumber \\
&+&2\left\vert \eta \right\vert ^{2}\left[ \left( \mathrm{i}\partial ^{\ast
}l+\langle a\rangle _{t}\right) \mathrm{d}W(t)^{\ast }+\left( \mathrm{i}%
\partial l+\langle a^{\dagger }\rangle _{t}\right) \mathrm{d}W(t)\right] \,.
\label{eq:5.25}
\end{eqnarray}

With the initial condition (\ref{eq:5.14}) the solution of (\ref{eq:5.25})
remains quadratic in $\xi $ and $\xi ^{\ast }$. Indeed, let us write 
\begin{equation}
l(\xi ^{\ast },\xi ;t)=\mathrm{i}\bigl[\xi ^{\ast }\langle a\rangle _{t}+\xi
\langle a^{\dagger }\rangle _{t}\bigr]+\frac{1}{2}\,\bigl[\xi ^{\ast 2}\mu
(t)+\xi ^{2}\mu (t)^{\ast }\bigr]+\left\vert \xi \right\vert ^{2}\nu (t)\,,
\label{eq:5.26}
\end{equation}%
where $\nu (t)\geq 0$; the term independent of $\xi $ is lacking because of
normalization of $\varrho (t)$ and the linear term must have just the form
we have written because $\langle a\rangle _{t}$ is the a posteriori mean
value of $\dot{W}(t)$. By inserting (\ref{eq:5.26}) into (\ref{eq:5.25}) and
equating the coefficients of the same order in $\xi $ and $\xi ^{\ast }$, we
obtain 
\begin{eqnarray}
\lefteqn{\mathrm{d}\langle a\rangle _{t}+\left[ \left( \mathrm{i}\omega +%
\frac{1}{2}\,\Gamma \right) \langle a\rangle _{t}+\mathrm{i}g(t)\right] 
\mathrm{d}t=}  \nonumber \\
&&2\left\vert \eta \right\vert ^{2}\left\{ \mu (t)\left[ \mathrm{d}%
W(t)^{\ast }-\langle a^{\dagger }\rangle _{t}\mathrm{d}t\right] +\nu (t)%
\left[ \mathrm{d}W(t)-\langle a\rangle _{t}\mathrm{d}t\right] \right\} \,,
\label{eq:5.27}
\end{eqnarray}%
\begin{equation}
{\frac{\mathrm{d}}{\mathrm{d}t}}\mu (t)+\left( 2\mathrm{i}\omega +\Gamma
\right) \mu (t)=-4\left\vert \eta \right\vert ^{2}\mu (t)\nu (t)\,,
\label{eq:5.28}
\end{equation}%
\begin{equation}
{\frac{\mathrm{d}}{\mathrm{d}t}}\nu (t)+\Gamma \nu (t)=-2\left\vert \eta
\right\vert ^{2}\left( \left\vert \mu (t)\right\vert ^{2}+\nu (t)^{2}\right)
+2\lambda _{\uparrow }\,,  \label{eq:5.29}
\end{equation}%
with $\langle a^{\dagger }\rangle _{t}=\langle a\rangle _{t}^{\ast }$ and
the initial conditions $\langle a\rangle _{0}=\alpha _{0}$, $\mu (0)=\mu
_{0} $, $\nu (0)=\nu _{0}$.

In the case $\mu_0 = 0$, we obtain $\mu(t) = 0$ $\big($the stationary
solution of (\ref{eq:5.28})$\big)$ and (\ref{eq:5.29}) becomes 
\begin{equation}
{\frac{\mathrm{d} }{\mathrm{d} t}} \nu(t) + \Gamma \nu(t) = -2
\left|\eta\right|^2 \nu(t)^2 +2 \lambda_\uparrow\,,  \label{eq:5.30}
\end{equation}
which is Riccati's equation and has the stationary positive solution $%
\nu_\infty$ 
\begin{equation}
\nu_\infty = {\frac{\Gamma }{4 \left|\eta\right|^2}} \left[ \left( 1 + 16
\left|\eta\right|^2 {\frac{\lambda_\uparrow }{\Gamma^2}} \right)^{1/2} - 1 %
\right]\,.  \label{eq:5.31}
\end{equation}
Equations (\ref{eq:5.27}) (for $\mu = 0$) and (\ref{eq:5.30}) were obtained
for the first time in \cite{bib:ref2,bib:XP31,bib:ref11} as optimal
filtering equations for linear systems.

After a transient any memory of the initial condition is lost. The
characteristic functional is given by (\ref{eq:5.17}) and (\ref{eq:5.18})
with a priori mean values 
\begin{equation}
\alpha (t)=-\mathrm{i}\int_{0}^{t}\mathrm{e}^{-(\mathrm{i}\omega +\Gamma
/2)(t-s)}\,g(s)\,\mathrm{d}s  \label{eq:5.32}
\end{equation}%
and covariance $\Delta _{2}(s,s^{\prime })=0$, 
\begin{equation}
\Delta _{1}(s,s^{\prime })={\frac{1}{2\left\vert \eta \right\vert ^{2}}}%
\delta (s-s^{\prime })+{\frac{2\lambda _{\uparrow }}{\Gamma }}\,\mathrm{e}%
^{-(\Gamma /2)\left\vert s-s^{\prime }\right\vert }\,\mathrm{e}^{-\mathrm{i}%
\omega (s-s^{\prime })}\,.  \label{eq:5.33}
\end{equation}%
The a priori states are given by 
\begin{equation}
\tilde{\sigma}_{\infty }(\xi ^{\ast },\xi ;t)=\exp \left\{ -\mathrm{i}\left[
\xi ^{\ast }\alpha (t)+\xi \alpha (t)^{\ast }\right] -{\frac{2\lambda
_{\uparrow }}{\Gamma }}\left\vert \xi \right\vert ^{2}\right\} \,,
\label{eq:5.34}
\end{equation}%
while the a posteriori states are 
\begin{equation}
\tilde{\varrho}_{\infty }(\xi ^{\ast },\xi ;t)=\exp \left\{ -\mathrm{i}\left[
\xi ^{\ast }\langle a\rangle _{t}+\xi \langle a^{\dagger }\rangle _{t}\right]
-\nu _{\infty }\left\vert \xi \right\vert ^{2}\right\} \,,  \label{eq:5.35}
\end{equation}%
with a posteriori mean values 
\begin{eqnarray}
\langle a\rangle _{t} &=&\int_{0}^{t}\exp \left[ -\left( \mathrm{i}\omega +%
\frac{1}{2}\,\Gamma +2\left\vert \eta \right\vert ^{2}\nu _{\infty }\right)
\left( t-s\right) \right]  \nonumber \\
&&\left[ -\mathrm{i}g(s)\,\mathrm{d}s+2\left\vert \eta \right\vert ^{2}\nu
_{\infty }\,\mathrm{d}W(s)\right] .  \label{eq:5.36}
\end{eqnarray}%
Note that $\nu _{\infty }>2\lambda _{\uparrow }/\Gamma $ for $\lambda
_{\uparrow }>0$ and $\nu _{\infty }=0$ for $\lambda _{\uparrow }=0$. In this
last case the asymptotic a priori and a posteriori mean values coincides $%
\big(\alpha (t)=\langle a\rangle _{t}\big)$ and the same holds for a priori
and a posteriori states 
\begin{equation}
\sigma _{\infty }(t)=\varrho _{\infty }(t)=|\alpha (t)\rangle \langle \alpha
(t)|\,  \label{eq:5.37}
\end{equation}%
where $|\alpha \rangle $ denotes the usual coherent states and $\alpha (t)$
is given by (\ref{eq:5.32}).


\end{document}